\documentclass[journal]{IEEEtran}
\usepackage{multirow}
\usepackage{booktabs}
\usepackage{makecell}
\usepackage[numbers,sort&compress]{natbib}
\usepackage{amsmath}
\usepackage{array}
\usepackage{graphicx}
\usepackage{float}
\usepackage{fixltx2e}
\usepackage{amssymb}
\usepackage{pifont}
\usepackage[table,xcdraw]{xcolor}
\graphicspath{{./figures/}}
\usepackage{url}  
\usepackage[noend]{algpseudocode}
\usepackage{algorithmicx,algorithm}
\usepackage{epstopdf}
\usepackage{color}
\usepackage{bbding} 
\usepackage{pifont}
\usepackage{xcolor}
\usepackage{natbib}
\usepackage{threeparttable}
\usepackage{titlesec}
\ifCLASSOPTIONcompsoc
  \usepackage[caption=false,font=normalsize,labelfont=sf,textfont=sf]{subfig}
\else
  \usepackage[caption=false,font=footnotesize]{subfig}
\fi

\begin{document}

\title{Efficient Seismic Data Interpolation via Sparse Attention Transformer and Diffusion Model}

\author{Xiaoli~Wei, Chunxia~Zhang, ~\IEEEmembership{Member, ~IEEE}, Baisong Jiang, Anxiang Di, Deng Xiong, Jiangshe~Zhang, Mingming Gong, ~\IEEEmembership{Member, ~IEEE} 
\thanks{Corresponding author: Chunxia Zhang. E-mail: cxzhang@mail.xjtu.edu.cn.}
\thanks{Xiaoli Wei, Chunxia Zhang, Baisong Jiang, Anxiang Di, Jiangshe~Zhang are with the School of Mathematics and Statistics, Xi'an Jiaotong University, Xi'an, Shaanxi, 710049, China.}
\thanks{Deng Xiong is with the Geophysical Technology Research and Development Center, BGP, Zhuozhou, Hebei, 072751, China}
\thanks{Mingming Gong is with the School of
Mathematics and Statistics, Faculty of Science, The University of Melbourne, Melbourne, VIC 3010, Australia.}
\thanks{This research was supported by the National Natural Science Foundation of China (No. 12371512), the Science \& Technology Research and Development Project of CNPC (Grant No. 2021ZG03), and the Key Acquisition, Processing and Interpretation Techniques for Seismic Data Processing of
CNPC (Grant No. 01-03-2023)}
}

\markboth{
Journal of \LaTeX\ Class Files,~Vol.~14, No.~8, August~2015}%
{Shell \MakeLowercase{\textit{et al.}}: Bare Demo of IEEEtran.cls for IEEE Journals
}

\maketitle

\begin{abstract}
Seismic data interpolation is a critical pre-processing step for improving seismic imaging quality and remains a focus of academic innovation. To address the computational inefficiencies caused by extensive iterative resampling in current plug-and-play diffusion interpolation methods, we propose the diffusion-enhanced sparse attention transformer (Diff-spaformer), a novel deep learning framework. Our model integrates transformer architectures and diffusion models via a Seismic Prior Extraction Network (SPEN), which serves as a bridge module. Full-layer sparse multi-head attention and feed-forward propagation capture global information distributions, while the diffusion model provides robust prior guidance. To mitigate the computational burden of high-dimensional representations, self-attention is computed along the channel rather than the spatial dimension. We show that using negative squared Euclidean distance to compute sparse affinity matrices better suits seismic data modeling, enabling broader contribution from amplitude feature nodes. An adaptive ReLU function further discards low or irrelevant self-attention values. We conduct training within a single-stage optimization framework, requiring only a few reverse diffusion sampling steps during inference. Extensive experiments demonstrate improved interpolation fidelity and computational efficiency for both random and continuous missing data, offering a new paradigm for high-efficiency seismic data reconstruction under complex geological conditions.

\end{abstract}


\IEEEpeerreviewmaketitle

\section{Introduction}
\IEEEPARstart{S}{eismic} data acquired in exploration surveys often suffer from spatial irregularity due to economic constraints and environmental factors, while complete regular seismic records are mandatory to the foundation for successful imaging and inversion in industry \cite{trad2009five}. Interpolation serves as a critical step to enhance the resolution and signal coherence of migrated seismic images and reduce artifacts caused by missing data, as it is capable of reconstructing complete wavefields. 

Traditional methods infer missing values by exploring the mathematical or physical properties of seismic data, leveraging spatial relationships and variation patterns of known data points. Predictive filter-based approaches employ spatial correlations between adjacent traces to estimate missing values through adaptive filters optimized to minimize prediction errors \cite{spitz1991seismic},  \cite{wang2002seismic}, \cite{chen20215d}. Wave equation-based methods reconstruct missing seismic data by inverting subsurface velocity models or wavefield information from available data, followed by forward simulations to generate complete wavefields that fill data gaps through physics-driven interpolation \cite{ronen1987wave}, \cite{fomel2002applications}. Transform-based methods project missing seismic data into mathematical transformation domains (e.g., 1D or multidimensional Fourier transform \cite{sacchi1998interpolation, zwartjes2007fourier1}, \cite{zwartjes2007fourier2}, sparsity-enhanced Curvelet transform \cite{naghizadeh2010beyond},  \cite{shahidi2013application}, and Radon transform \cite{wang2010seismic}) and exploit sparse signal priors to recover gaps through domain-specific optimization. Rank-reduction methods interpolate seismic data by leveraging low-rank structures and linear correlations in transform domains, recovering missing traces via matrix completion or nuclear norm optimization \cite{Trickett2010Rank, ma2013three}. 

The above traditional methods rely on rigid assumptions and prior knowledge, limiting their adaptability. Data-driven approaches leverage adaptive learning to bypass these constraints, emerging as popular solutions in contemporary seismic interpolation. The symmetric encoder-decoder architecture of U-Net establishes a stable mapping mechanism for seismic data processing \cite{Xinming101190}, \cite{wang2025quadratic}, \cite{Hongtao10630534}, making it a dominant framework for interpolation tasks \cite{mandelli2018seismic}, \cite{meng2021self}, \cite{tang2020reconstruction}. Recent U-Net enhancements integrate physics-aware loss functions \cite{fang2021seismic}, partial convolution layers (adaptive feature focusing via validity-guided attention) \cite{pan2020partial}, and wavelet-based downsampling (preserving high-frequency components) \cite{liu2022seismic}. GAN-based methods improve seismic interpolation via adversarial learning \cite{kaur2019seismic}, \cite{wei2022big}, where discriminator-guided training drives the generator toward data distribution alignment, enhancing geological fidelity. 
Multi-branch \cite{Abedi10531775} and multi-stage interpolation methods \cite{He9469034}, \cite{WANGBenFeng2024}, \cite{song9893067} hierarchically integrate diverse feature extraction paths with stepwise optimization, enhancing accuracy through adaptive fusion of seismic attributes and phased refinement. Dual-domain conditional generative adversarial network (DD-CGAN) \cite{Chang2020} utilizes joint supervision of spatiotemporal and time-frequency domains in seismic data feature restoration, balancing energy distribution optimization in the time-frequency domain with waveform variation preservation in the spatiotemporal domain. The Coarse-to-Fine model \cite{wei2022hybrid} employs depth-varied feature extractors and differentiated supervision for progressive low/high amplitude interpolation. Attention-based methods improve feature saliency by capturing global context and long-range dependencies \cite{Yu9390348}, \cite{rs16020305}, \cite{Bangyu9837951}, addressing the locality limitation of convolutions, which is crucial for reconstructing continuous missing seismic data. The transformer enhances this advantage with multi-head attention \cite{Guo10190759}, enabling parallel multi-scale seismic feature interactions. Physics-informed and physics-guided models significantly enhance solution stability in data-driven frameworks through convex constraint enforcement. One of the most well-known convex optimization frameworks, projection onto convex sets (POCS), has been extensively studied and combined with deep neural networks to solve interpolation inverse problems \cite{zhang2020can}, \cite{chen2023projection}. Physics-Informed Neural Networks (PINN) offer a new paradigm that explicitly constrains the continuity of the interpolation process using plane wave differential equations, promoting reconstructions naturally similar to the available data \cite{RAISSI2019686}, \cite{brandolin2024pinnslope}.
 
Recently, diffusion models for interpolation have become a research focus \cite{durall2023deep}, \cite{liu2024generative}, bridging deep learning with iterative frameworks. Using Langevin dynamics and a predefined variance schedule, missing seismic data is mapped into a noisy distribution flow, enabling a deep neural network to learn nonlinear mappings across noise levels. However, optimizing valid data usage in reverse conditional interpolation remains a key challenge. While resampling and conditional generation strategies have shown promise in the prior study \cite{Lugmayr_2022_CVPR}, the plug-and-play approach falls short in complex seismic data reconstruction. Subsequent work has thus incorporated self-correction mechanisms \cite{10731876},  posterior distribution through Langevin dynamics \cite{Meng10579850}, and constrained training paradigms \cite{deng2024seismic}, \cite{10681481}, \cite{Xinlei10440113} to enhance interpolation accuracy and stability. However, these methods inevitably face a common challenge: prolonged inference time due to multiple sampling iterations. Furthermore, the presence of various high-performance feature extractors, especially the remarkable ability of multi-head attention mechanisms to effectively model and manage both random and continuous missing data scenarios, deserves significant attention. This insight motivates the integration of transformer architectures with diffusion models. In this paper, we propose the diffusion-enhanced sparse attention transformer (Diff-spaformer), which synergistically combines multi-scale transformer networks for efficient long-sequence modeling with a diffusion-based prior generator to maintain distribution consistency, enabling efficient interpolation of missing seismic data. Our method outperforms some existing diffusion-based approaches in both computational efficiency and accuracy. The main contributions are summarized below.

\begin{itemize}
\item We propose a novel framework that integrates transformer networks with diffusion models through a bridge component called the Seismic Prior Extraction Network (SPEN). 
Full-layer sparse multi-head attention and feed-forward propagation synergistically capture global information distributions, while the diffusion model provides strong prior guidance to enhance reconstruction quality.  

\item To reduce the computational complexity of high-dimensional space representations, we reorient the self-attention computation from the spatial dimension to the channel dimension. We adopt the negative squared Euclidean distance as similarity functions to compute this sparse affinity matrix, facilitating more feature nodes to make contributions. The improved sparse transformer block can model the global pixel relationships with higher efficiency and effectiveness. 

\item  We use a simple yet effective ReLU function as the activation function to adaptively remove the self-attention values with low/no correlation. 

\item Extensive experimental results show that only adopting a single-stage unified multi-component optimization, our method can deliver superior performance while avoiding excessive computational overhead.

\end{itemize}
The content of this paper is outlined as follows. Section \ref{Background} introduces the background. Section \ref{secMethodology} proposes the methodological framework and describes the detailed components. Section \ref{secExperiments} presents experimental results. In Section \ref{secabalation}, we have conducted ablation studies, and Section \ref{secConclusion} concludes this paper.

\section{Background}\label{Background}
\begin{figure*}[!htbp]
	\centering
    \includegraphics[width=6.8in]{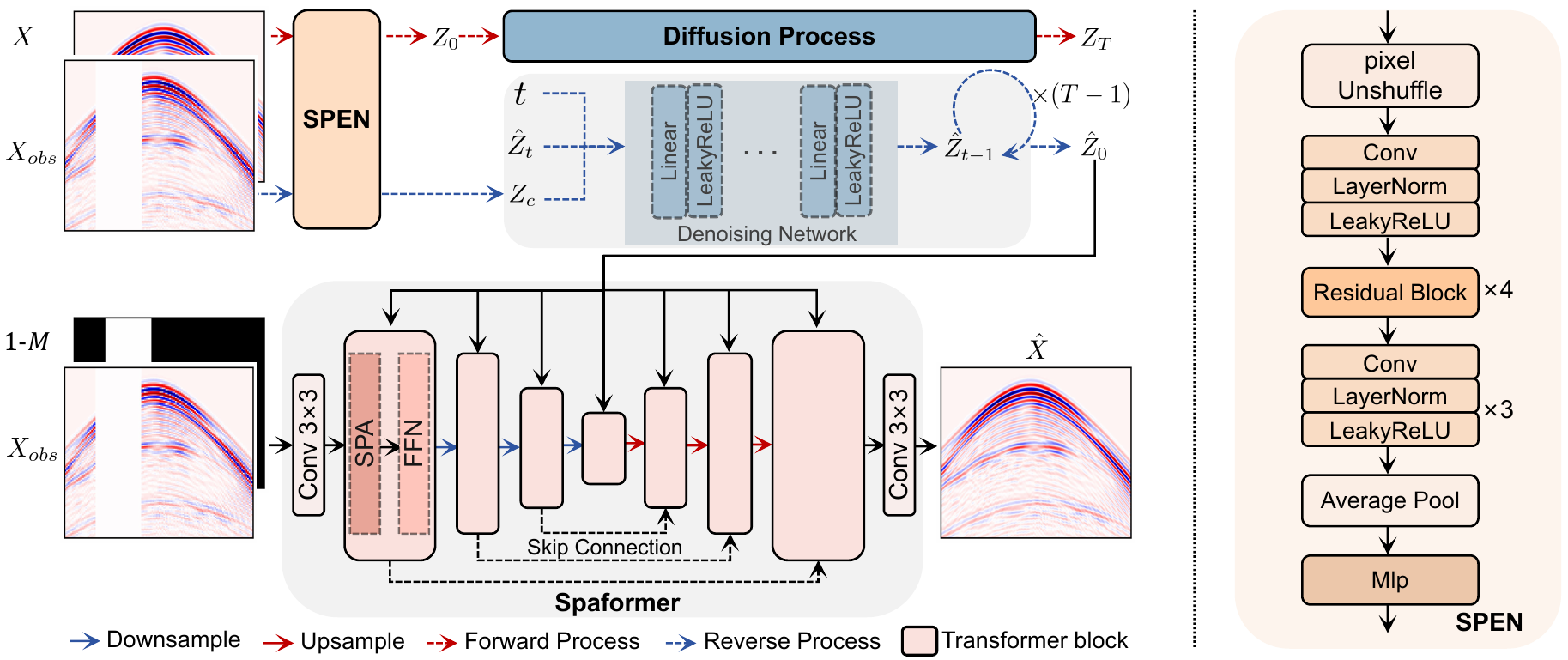}
    \vspace{-2mm}
	\caption{The overall framework of Diff-spaformer. It comprises three components, i.e., SPEN, diffusion process, and Spaformer. SPEN encodes seismic prior information and generates latent features through the diffusion model. The Spaformer module receives the concatenation of the missing data $\boldsymbol{X}_{obs}$ and the missing mask $1$-$M$ as the input while integrating seismic prior knowledge extracted from SPEN at each encoding-decoding layer. This hierarchical fusion mechanism ultimately outputs the interpolated data $\hat{\boldsymbol{X}}$.}
	\label{fig:diffspaformer}
\vspace{-3mm}
\end{figure*}

Denoising Diffusion Probabilistic Models (DDPM), as defined in \cite{ho2020denoising}, achieve complex distribution modeling by learning a progressive noise-adding and denoising process of data distributions. The progressive noising process constitutes a predefined $T$-step forward process, and the denoising process executes a $T$-step reverse process to construct desired data samples from noises. 
\subsection{Forward Process}\label{Forward Process}
In the forward process, the noise-adding operations between adjacent timesteps are designed as a Markov chain $\boldsymbol{x}_0 \rightarrow \boldsymbol{x}_{1} \rightarrow \ldots \rightarrow \boldsymbol{x}_{T-1} \rightarrow \boldsymbol{x}_T\left(\right.$where $\left.\boldsymbol{x}_t \in \boldsymbol{R}^n\right)$. 
The state transition from the previous state $\boldsymbol{x}_{t-1}$ to the current state $\boldsymbol{x}_{t}$ is mathematically defined by a conditional Gaussian distribution parameterized with scheduled variance coefficients $\beta_t$ as
\begin{equation}
\label{equ:q_t_t1}
q\left(\boldsymbol{x}_t \mid \boldsymbol{x}_{t-1}\right):=\mathcal{N}\left(\boldsymbol{x}_t ; \sqrt{1-\beta_t} \boldsymbol{x}_{t-1}, \beta_t \mathbf{I}\right),
\end{equation}
i.e., 
\begin{equation}
\label{equ:q_t_t1_add}
\boldsymbol{x}_t=\sqrt{1-\beta_t} \boldsymbol{x}_{t-1}+\sqrt{\beta_{t}} \boldsymbol{\epsilon}, \quad \boldsymbol{\epsilon} \sim \mathcal{N}(0, \mathbf{I}).
\end{equation}
The reparameterization technique integrates these multi-step stochastic processes into the following closed-form expression as
\begin{equation}
\label{equ:q_t}
q\left(\boldsymbol{x}_t \mid \boldsymbol{x}_0\right)=\mathcal{N}\left(\boldsymbol{x}_t ; \sqrt{\bar{\alpha}_t} \boldsymbol{x}_0,\left(1-\bar{\alpha}_t\right) \mathbf{I}\right),
\end{equation}
where $\alpha_t=1-\beta_t$ and $\bar{\alpha}_t=\prod_{i=0}^t \alpha_i$.

\subsection{Reverse Process}\label{Reverse Process}
Each step in the reverse process achieves state transition through a conditional probability distribution $p\left(\boldsymbol{x}_{t-1} \mid \boldsymbol{x}_t, \boldsymbol{x}_0\right)$. According to Bayes theorem, it can be derived from Eq. (\ref{equ:q_t_t1}) and Eq. (\ref{equ:q_t}) as
\begin{equation}
\label{equ:p_t_t-1}
p\left(\boldsymbol{x}_{t-1} \mid \boldsymbol{x}_t, \boldsymbol{x}_0\right)=\mathcal{N}\left(\boldsymbol{x}_{t-1} ; \boldsymbol{\mu}_t\left(\boldsymbol{x}_t, \boldsymbol{x}_0\right), \sigma_t^2 \mathbf{I}\right),
\end{equation}
where 
\begin{equation*}
\boldsymbol{\mu}_t\left(\boldsymbol{x}_t, \boldsymbol{x}_0\right)=\frac{1}{\sqrt{\alpha_t}}\left(\boldsymbol{x}_t-\frac{1-\alpha_t}{\sqrt{1-\bar{\alpha}_t}}\boldsymbol{\epsilon}_t \right)
\end{equation*}
and 
\begin{equation*}
\sigma_t^2=\frac{1-\bar{\alpha}_{t-1}}{1-\bar{\alpha}_t} \beta_t.
\end{equation*}
The $\theta$-parameterized neural network estimates the noise component $\boldsymbol{\epsilon}_t$ to recover the noiseless observation from the noisy input $\boldsymbol{x}_t$, i.e., $\boldsymbol{\epsilon}_t = \boldsymbol{\epsilon}_\theta\left(\boldsymbol{x}_t, t\right)$.
The Gaussian form of the target distribution in Eq. (\ref{equ:p_t_t-1}) simplifies the measurement of the Kullback-Leibler (KL) divergence, enabling efficient variational training through the optimization of the evidence lower bound (ELBO). The simplified training objective used in DDPM is formulated as 
\begin{equation}
\label{equ:elbo}
\mathbb{E}_{\boldsymbol{x}_0, t, \epsilon}\left[\left\|\epsilon-\boldsymbol{\epsilon}_\theta\left(\sqrt{\bar{\alpha}_t} \boldsymbol{x}_0+ \sqrt{1-\bar{\alpha}_t}\boldsymbol{\epsilon}, t\right)\right\|_2^2\right].
\end{equation}
where $\|\cdot\|_2$ symbolizes L2 norm and $t$ is uniformly distributed over the interval from $1$ to $T$.

\section{Methodology}\label{secMethodology}
Given the complete seismic data denoted as $\boldsymbol{X} \in \mathbb{R}^{n_{s} \times n_{t}}$ (where $n_{s}$ and $n_{t}$ represent the spatial and  temporal dimensions, respectively) and the degraded observation data $\boldsymbol{X}_{obs}$ containing missing traces, the interpolation process is typically modeled as an inverse problem:
\begin{equation}
\label{equ:observation_form_complete}
\boldsymbol{X}_{obs} = \boldsymbol{M} \odot \boldsymbol{X}, 
\quad 
\text{with} \quad \boldsymbol{M}[i,:] = 
\begin{cases} 
\boldsymbol{1}, & \text{if } i \text{ is valid} \\
\boldsymbol{0}, & \text{ else }
\end{cases}
\end{equation}
where $\odot$ represents the element-wise multiplication and $\boldsymbol{M}$ denotes the binary mask matrix with continuous or random missing traces. Our proposed Diff-spaformer model aims to learn the nonlinear relationship between degraded observations and fully sampled seismic data through the parameterized mapping process: 
\begin{equation}
\hat{\boldsymbol{X}} =  f_\theta \left( \boldsymbol{X}_{obs}, \boldsymbol{Z}\right), 
\end{equation}
where $\boldsymbol{Z}$ represents the seismic prior feature. Fig. \ref{fig:diffspaformer} demonstrates the specific workflow of our proposed Diff-spaformer model. Spaformer is built on the U-Net architecture, equipped with transformer modules, and the diffusion process further provides compressed prior information to guide Spaformer. SPEN serves as a bridge between Spaformer and the diffusion process. Section \ref{SPEN}, Section \ref{Diffusion Process}, and Section \ref{Spaformer} will introduce the details of the three parts in turn.

\subsection{Seismic Prior Extraction Network (SPEN)}\label{SPEN}
SPEN operates externally to the Spaformer architecture, focusing on capturing inherent seismic characteristics like amplitude patterns and spectral properties. This auxiliary module supplements the primary encoder-decoder framework by integrating domain-specific structural constraints, thereby optimizing reconstruction accuracy through enhanced distribution modeling of seismic signals.
We adopt the prior network design used in previous restoration works \cite{Xia_2023_ICCV}, as shown on the right side of Fig. \ref{fig:diffspaformer}. Given the inherent high spatial continuity of seismic data properties (e.g., amplitude structure and frequency distribution), we implement spatial-to-depth transformation through PixelUnshuffle downsampling. This operation effectively preserves complete signal energy while extracting multi-scale features through spatial dimension reorganization, circumventing detail loss inherent in conventional pooling methods through channel expansion rather than spatial compression. 
The network employs a convolutional-residual structure for stable feature extraction, where convolutional layers capture local spatial correlations and residual layers enhance gradient propagation stability through cross-layer connections. Then, average pooling is applied to the feature maps for spatial compression, reducing dimensionality while retaining global statistical features and suppressing local noise. Finally, a multilayer perceptron (MLP) performs nonlinear transformations and high-order feature fusion on the compressed features. The SPEN generates compressed prior information $\boldsymbol{Z}_0 \in \mathbb{R}^{C^{\prime}}$ that represents the energy distribution of seismic signals as 
\begin{equation}
\label{equ:spen_x}
\boldsymbol{Z}_0 = \operatorname{SPEN} \left( \boldsymbol{X}\right).
\end{equation}
At the same time, conditional prior information $\boldsymbol{Z}_c \in \mathbb{R}^{C^{\prime}}$ for the diffusion process is provided by encoding features from the observed data as follows:
\begin{equation}
\label{equ:spen_obs}
\boldsymbol{Z}_c = \operatorname{SPEN} \left( \boldsymbol{X}_{obs}\right).
\end{equation}
In diffusion model training, the original data prior $\boldsymbol{Z}_0$ and conditional prior $\boldsymbol{Z}_c$ are critical for optimizing the iterative process. The original distribution captures the intrinsic probability characteristics of the source data, while the conditional prior encodes domain-specific constraints or observed patterns from incomplete datasets through feature extraction mechanisms. This dual-input way enables the model to learn conditional distributions guided by known information during optimization.

\subsection{Diffusion Process}\label{Diffusion Process}
We exploit the $T$-steps diffusion forward process $\boldsymbol{Z}_0 \rightarrow \boldsymbol{Z}_{1} \rightarrow \ldots \rightarrow \boldsymbol{Z}_{T-1} \rightarrow \boldsymbol{Z}_T$ by executing the predefined progressive noise-adding operation
\begin{equation}
\label{equ:q_t_z}
q\left(\boldsymbol{Z}_t \mid \boldsymbol{Z}_0\right)=\mathcal{N}\left(\boldsymbol{Z}_t ; \sqrt{\bar{\alpha}_t} \boldsymbol{Z}_0,\left(1-\bar{\alpha}_t\right) \mathbf{I}\right),
\end{equation}
where $\bar{\alpha}_t$ has been defined in Eq. (\ref{equ:q_t}) and $\beta_t \in(0,1)$ follows a linear increasing schedule. The reverse process starts from $T$-th time step and $\boldsymbol{Z}_T$ is randomly sampled from the standard Gaussian distribution. Since the seismic priors $\boldsymbol{Z}_0$ have been encoded by SPEN into compact features, the complex distribution of the original seismic data is effectively projected into a low-dimensional latent space. These latent space features serve as the strong valid priors, effectively guiding the diffusion model's training and allowing us to capture key noise patterns using a simple and small network without complex architectures. As shown in Fig. \ref{fig:diffspaformer}, this noise-matching network $\boldsymbol{\epsilon}_\theta$ is only composed of multiple linear layers. During the reverse process, it uses the concatenated inputs of $t$, $\hat{\boldsymbol{Z}}_t$, and $\boldsymbol{Z}_c$ to directly predict the raw data corresponding to the current time step $t$ as
\begin{equation}
\label{equ:network_eps_pred}
\hat{\boldsymbol{Z}}_{0}^{t}= \epsilon_\theta\left(\left[t, \hat{\boldsymbol{Z}}_t, \boldsymbol{Z}_c\right]\right),
\end{equation}
where $t$ is embedded by using $t=\frac{t}{T}$. The final compact prior feature $\hat{\boldsymbol{Z}_0}$ can be obtained by performing iterative generation
\begin{equation}
\label{equ:sampling_z_pred}
\hat{\boldsymbol{Z}}_{t-1}=\frac{\sqrt{\bar{\alpha}_{t-1}} \beta_t}{1-\bar{\alpha}_t} \hat{\boldsymbol{Z}}_{0}^{t}+\frac{\sqrt{\alpha_t}\left(1-\bar{\alpha}_{t-1}\right)}{1-\bar{\alpha}_t} \hat{\boldsymbol{Z}}_t,
\end{equation}
where the noise term is excluded by reducing stochasticity to enhance generation efficiency and stability, particularly in our scenarios with simpler model architectures and smaller scales. Due to the long sampling time $T$ in traditional diffusion model training, loss function computation and gradient backpropagation are typically carried out using randomly sampled sequences. However, with the smaller and simpler structure in our case, we can compute the loss over all time steps during each training iteration as 
\begin{equation}
\label{equ:loss_diff}
\mathcal{L}_{\text {diff}}=\frac{1}{T-1} \sum_{t=0}^{T-1 }\|\boldsymbol{Z}_t-\hat{\boldsymbol{Z}}_t\|_1,
\end{equation}
where $\|\cdot\|_1$ represents the L1 norm, facilitating the optimization to obtain an even sharper distribution. 

In the inference phase, SPEN directly encodes missing data $\boldsymbol{X}_{obs}$ to produce conditional latent features $\boldsymbol{Z}_c$, subsequently generating prior latent features $\hat{\boldsymbol{Z}}_0$ via the noise matching and diffusion sampling in Eqs. (\ref{equ:network_eps_pred}) and (\ref{equ:sampling_z_pred}).

\subsection{Spaformer}\label{Spaformer}
The Spaformer module serves as the core component of our Diff-spaformer model to establish an end-to-end network mapping from missing data to complete data. As Fig. \ref{fig:diffspaformer} shows, the concatenated input of observed seismic data $\boldsymbol{X}_{obs}$ and its corresponding missing mask $1$-$M$ is first processed by a $3\times3$ convolutional layer for channel dimension alignment. 
The encoding-decoding hierarchy utilizes U-Net's structural paradigm enhanced with transformer blocks, where each block integrates a sparse attention (SPA) mechanism for contextual feature modeling and a feedforward network (FFN) for feature transformation and enhancement. 
This hierarchical architecture balances signal fidelity and feature expression efficiency through progressive processing from global to local. Crucially, seismic prior knowledge is systematically integrated through cross-layer feature fusion mechanisms, ensuring consistent domain-specific constraint enforcement throughout the network. The final reconstruction phase employs a $3\times3$ convolutional layer to generate interpolated seismic data outputs $\hat{\boldsymbol{X}}$ while maintaining dimensional consistency with the input space. Specifically, suppose the input feature and output feature are $\boldsymbol{F}$ and $\hat{\boldsymbol{F}}$, respectively, we combine the SPA module with the FFN module using the following residual connection:
\begin{equation}
\begin{aligned}
\hat{\boldsymbol{F}} = \operatorname{FFN} \left(\boldsymbol{F}^{\prime},\hat{\boldsymbol{Z}}_0\right)+\boldsymbol{F}^{\prime},
\text{ where }\boldsymbol{F}^{\prime}=\operatorname{SPA} \left( \boldsymbol{F},\hat{\boldsymbol{Z}}_0\right)+\boldsymbol{F}.
\end{aligned}
\end{equation}
The SPA module processes the normalized input feature $\boldsymbol{F}$ along with the prior latent feature $\hat{\boldsymbol{Z}}_0$, and adds its output to the original input $\boldsymbol{F}$ via a residual connection to obtain $\boldsymbol{F}^{\prime}$. Then, the normalized feature $\boldsymbol{F}^{\prime}$, combined with the prior latent feature $\hat{\boldsymbol{Z}}_0$, serves as the input to the FFN module. The FFN output is added back to its input through another residual connection to generate $\hat{\boldsymbol{F}}$. This dual residual structure preserves the original feature information and enables the network to learn the residual mapping, effectively mitigating the vanishing gradient problem.

During model optimization, the L1 norm serves as the primary term in the loss function. To implement differentiated error weighting between valid traces and missing traces, we develop the weighted composite loss framework as
\begin{equation}
\label{equ:loss_rec}
\mathcal{L}_{\text {rec}}=\lambda_1\| \left(1-\boldsymbol{M}\right) \odot (\hat{\boldsymbol{X}}-\boldsymbol{X})\|_1 + \lambda_2 \| \boldsymbol{M} \odot (\hat{\boldsymbol{X}}-\boldsymbol{X})\|_1, 
\end{equation}
where $\lambda_1,\lambda_2 \geq 0$ to regulate the proportional weights of different loss components. This proposed formulation achieves higher error sensitivity in missing regions and enhances the model's capacity to detect and respond to discrepancies within data gaps through dynamically adjusted penalty mechanisms, while maintaining balanced optimization for valid signal regions. The final loss formulation can be expressed as
\begin{equation}
\label{equ:loss_total}
\mathcal{L}=\mathcal{L}_{\text {rec}}+\lambda_3\mathcal{L}_{\text {diff}}, 
\end{equation}
where the coefficient $\lambda_3 \geq 0$. This ensures that the optimization process prioritizes specific error terms according to their assigned weights and balances the trade-off between different objectives.
\begin{figure*}[!htbp]
  \centering
  {\includegraphics[width=0.9\textwidth]{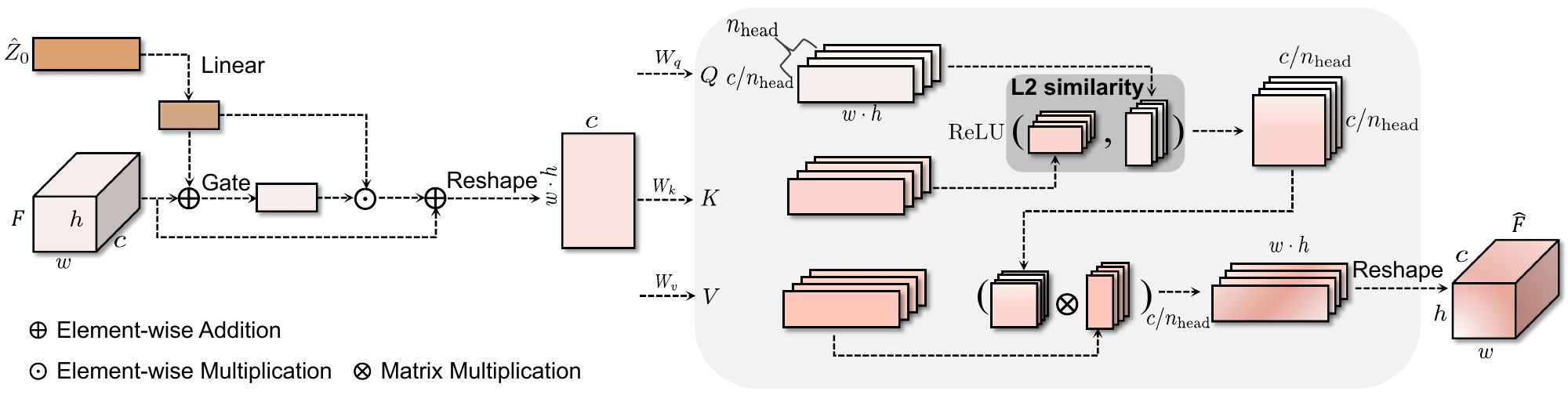}}
  \vspace{-2mm}
  \caption{SPA module}
  \label{fig:SPA module}
\vspace{-3mm}
\end{figure*}

\subsection{SPA Module}\label{SPA}
The SPA module integrates the compact seismic prior information extracted from SPEN, achieving multi-scale feature modeling from global to local scales. This process incorporates several key improvements, including dynamic feature calibration mechanisms, sparse self-attention, and affinity matrix computation based on negative squared Euclidean distance. We will systematically elaborate on these modules.

\subsubsection{Dynamic Feature Calibration Mechanism} \label{sec:Dynamic Feature Calibration Mechanism}
As Fig. \ref{fig:SPA module} shows, the dynamic feature calibration mechanism enables the adaptive fusion of external prior information $\hat{\boldsymbol{Z}}_0$ with the main feature flow. Let $\boldsymbol{F} \in \mathbb{R}^{c \times h \times w}$ be the input feature ($c$ is the channel dimension, and $h$ and $w$ represent spatial dimensions). As Fig. \ref{fig:SPA module} shows, we perform feature alignment on the seismic prior $\hat{\boldsymbol{Z}}_0$ through the linear transformation as 
\begin{equation}
\hat{\boldsymbol{Z}} =\boldsymbol{W}_z \cdot \hat{\boldsymbol{Z}}_0,  
\end{equation}
where $\boldsymbol{W}_z \in \mathbb{R}^{c\times C^{\prime}}$ is the weight matrix. After reshaping $\hat{\boldsymbol{Z}}$, the channel attention mechanism is applied to generate gating weights:
\begin{equation}
\boldsymbol{G}=\sigma\left(\operatorname{Conv}_2\left(\operatorname{ReLU}\left(\operatorname{Conv}_1\left(\operatorname{AvgPool}\left(\boldsymbol{F}+\hat{\boldsymbol{Z}}\right)\right)\right)\right)\right),
\end{equation}
where $\sigma$ denotes a Sigmoid activation function. Then, we adopt the following gated residual connections to achieve feature fusion:
\begin{equation}
\hat{\boldsymbol{F}}_G=\boldsymbol{F} + \boldsymbol{G} \odot \hat{\boldsymbol{Z}},
\end{equation}
where $\odot$ denotes the Hadamard product, and this operation allows the network to dynamically adjust the contribution of the seismic prior. Compared to traditional methods such as direct addition or channel concatenation, the dynamic feature calibration mechanism can more effectively realize adaptive fusion of cross-modal features. 

\subsubsection{Sparse Self-attention} 
We divide the input feature $\hat{\boldsymbol{F}}_G$ into different subspaces ($n_{\text{head}}$ in total) through multiple sets of independent linear projections as 
\begin{equation}
\boldsymbol{Q}_i=\hat{\boldsymbol{F}}_G \boldsymbol{W}_i^Q, \boldsymbol{K}_i=\hat{\boldsymbol{F}}_G \boldsymbol{W}_i^K, \boldsymbol{V}_i=\hat{\boldsymbol{F}}_G \boldsymbol{W}_i^V, i=1,\ldots,n_{\text{head}}
\end{equation}
where the trainable parameter matrices $\boldsymbol{W}_i^Q,\boldsymbol{W}_i^K,\boldsymbol{W}_i^V \in \mathbb{R}^{c\times c_n}\left(c_n = c/n_{\text{head}}\right)$.
Then, the multi-head self-attention \cite{NIPS2017_3f5ee243} can be constructed by 
\begin{equation}
\begin{aligned}
\operatorname{MultiHead}(\boldsymbol{Q}, \boldsymbol{K}, \boldsymbol{V}) &=\operatorname{Concat}\left(\operatorname{head}_1, \ldots, \operatorname{head}_{n_{\text{head}}}\right)\boldsymbol{W}^O, 
\\
\text {where } \operatorname{head}_i &=\operatorname{Attention}\left(\boldsymbol{Q}_i, \boldsymbol{K}_i, \boldsymbol{V}_i\right),
\end{aligned}
\end{equation}
where the output parameter matrix $\boldsymbol{W}^O\in \mathbb{R}^{c\times c}$ effectively integrates information from multiple heads to avoid redundancy.
Each head independently performs the attention calculation, allowing the model to simultaneously focus on different types of dependencies within the feature. It parallelizes multi-scale feature learning to enhance the model's ability to capture complex dependencies while maintaining computational efficiency.

Given $\boldsymbol{Q}_i,\boldsymbol{K}_i \in \mathbb{R}^{c_n\times N} \left(N=w\times h, \text{ usually } c_n \ll N \right)$ within the same head, the pairwise affinity matrix between them is $\boldsymbol{S}_{i}=s\left(\boldsymbol{Q}_i^T, \boldsymbol{K}_i\right)$,  typically $\boldsymbol{S}_{i} \in \mathbb{R}^{N\times N}$. This operation gives rise to $\mathcal{O}\left(N^2 c_n\right)$ computations and $\mathcal{O}\left(N^2\right)$ memory costs. Unlike general global attention mechanisms that solely operate on compressed features (e.g., ANet \cite{Yu9390348}), computational expenses become unacceptable when applied to the high-resolution features in the shallow layers of our model. To solve this issue, we reformulate attention mechanisms across channel dimensions. Specifically, global dependencies are captured through channel-wise dot-product operations between query $Q_i$ and key $K_i$ projections as
\begin{equation}
\label{equ:s_simality_ours}
\boldsymbol{S}_{i}=s\left(\boldsymbol{Q}_i, \boldsymbol{K}_i^T\right) \text{ with } \boldsymbol{S}_{i} \in \mathbb{R}^{c_n\times c_n}.
\end{equation}
The computational complexity and memory cost have been reduced to $\mathcal{O}\left(Nc_n^2\right)$ and $\mathcal{O}\left(c_n^2 \right)$, respectively, effectively improving the computational efficiency.

In conventional attention mechanisms, similarity scores are normalized using the softmax function, creating a dense weight matrix. While this enables dynamic attention allocation, it can lead to scattered weights, causing the model to attend to noise or weakly related elements, which impedes key information extraction. To address this issue, we adopt a sparse attention mechanism proposed in \cite{HUANG2024109897} as 
\begin{equation}
\label{equ:s_simality_ours_relu}
\boldsymbol{W}_{i}=\frac{1}{\omega} \operatorname{Re L U}\left(\boldsymbol{S}_{i}\right).
\end{equation}
where $\omega$ is a learnable scaling parameter. The Rectified Linear Unit (ReLU) activation function is  applied to conduct sparse similarity score calculation, and its nonlinear characteristics filter out negative values (low correlation or noise signals), retaining only positive values and directly suppressing weak correlations. This approach eliminates the reliance on softmax normalization and instead achieves sparsity through activation thresholding. The generated sparse attention map focuses on strongly related elements, reducing the interference of irrelevant information and lowering computational complexity.

Finally, the sparse attention map achieves feature-weighted aggregation by performing matrix multiplication with the Value matrix $\boldsymbol{V}_i$ per head. Specifically, the mathematical formulation for this process is: 
\begin{equation}
\label{equ:s_simality_ours_relu_final}
\text{head}_{i}= \boldsymbol{W}_i\boldsymbol{V}_i.
\end{equation}
We concatenate all heads and restore it to the original dimension to generate the multi-head feature $\operatorname{Concat}\left(\operatorname{head}_1, \ldots, \operatorname{head}_{n_{\text{head}}}\right)$ (i.e., $\hat{\boldsymbol{F}}$ in Fig. \ref{fig:SPA module}).



\subsubsection{Negative Squared Euclidean Distance}
The similarity definition in Eq. (\ref{equ:s_simality_ours}) is the core technological approach for constructing feature projections. Its essence lies in quantifying the strength of the association between feature vectors. Cosine similarity is a widely used metric that quantifies the similarity between vectors based on their directional alignment in the vector space. Its core principle involves evaluating the degree of similarity between two non-zero vectors by calculating the cosine of the angle between them. However, studies have shown that for tasks requiring strong global pixel attention, the optimization should focus on balancing both directional similarity and magnitude information rather than neglecting one in favor of the other \cite{cheng2021rethinking}.
In seismic data interpolation, accurate recovery of high-frequency details depends on magnitude information to represent structural similarity. The variation in high-frequency signal magnitude reflects strata heterogeneity and acoustic impedance differences. Thus, constructing the seismic similarity measure requires considering both the absolute magnitude and phase characteristics of the waveform. 

\begin{figure}[!htbp]
\centering
\subfloat[]{
    \includegraphics[height=4.0cm]{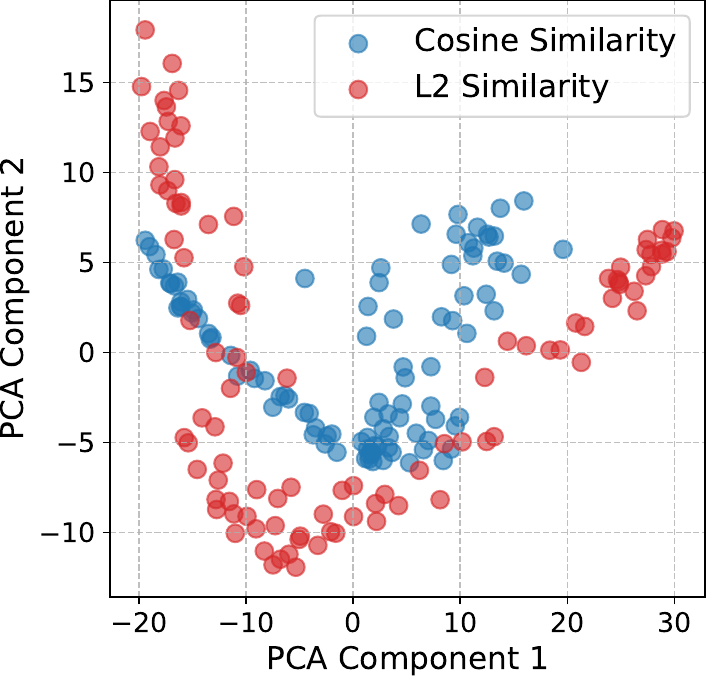}
    \label{fig:similarity_a}
}
\subfloat[]{
    \includegraphics[height=4.0cm]{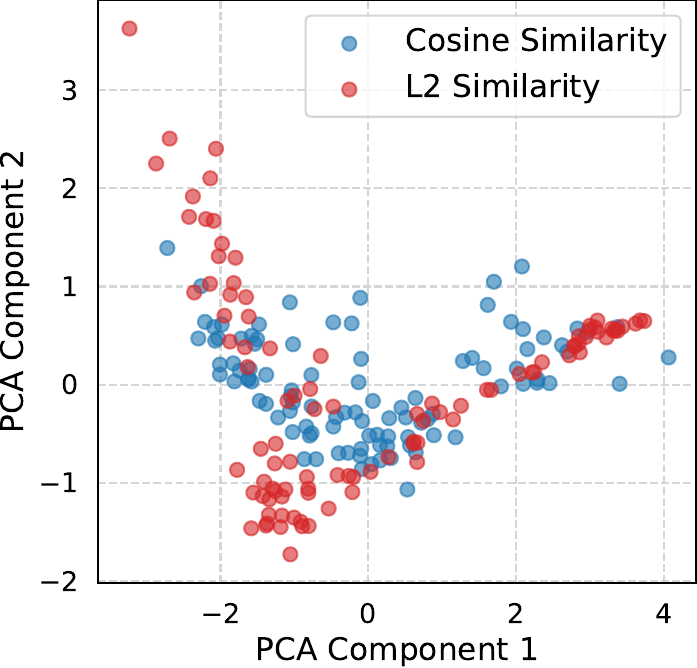}
    \label{fig:similarity_b}
}
\vspace{-2mm}
\caption{(a) PCA projection of the affinity matrix $\operatorname{Concat}\left(\boldsymbol{S}_1, \ldots, \boldsymbol{S}_{n_{\text{head}}}\right)$ under different similarity functions. (b) PCA projection of $\operatorname{MultiHead}(\boldsymbol{Q}, \boldsymbol{K}, \boldsymbol{V})$ under different similarity functions.}
\label{fig:similarity}
\vspace{-3mm}
\end{figure}

Let $\boldsymbol{S}_{i,jk}=s\left(\boldsymbol{Q}_{i,j}, \boldsymbol{K}_{i,k}^T\right)$
represent the similarity between the query feature vector $\boldsymbol{Q}_{i,j}$ (at the $j$-th position) and the key feature vector $\boldsymbol{K}_{i,k}^T$ (at the $k$-th position). The negative squared Euclidean distance for similarity functions is defined as
\begin{equation}
\label{equ:s_simality_l2}
s\left(\boldsymbol{Q}_{i,j}, \boldsymbol{K}_{i,k}^T\right)=-\left\|\boldsymbol{Q}_{i,j}-\boldsymbol{K}_{i,k}^T\right\|_2^2, 
\end{equation}
whose computational complexity is only slightly higher than that of cosine similarity. For simplicity, we refer to it as L2 similarity. Its core is to measure the Euclidean distance between vectors, converting the distance into a similarity score through a negative sign (the smaller the distance, the higher the score). Unlike the cosine similarity, L2 similarity is not dominated by the vector magnitudes, thus avoiding the interference of seismic amplitude differences on attention weights. 
Fig. \ref{fig:similarity}(a) and Fig. \ref{fig:similarity}(b) visualize the projections of the affinity matrix $\operatorname{Concat}\left(\boldsymbol{S}_1, \ldots, \boldsymbol{S}_{n_{\text{head}}}\right)$ and the output features $\operatorname{MultiHead}(\boldsymbol{Q}, \boldsymbol{K}, \boldsymbol{V}) $ from the multi-head attention module after dimensionality reduction using Principal Component Analysis (PCA). It can be observed that, compared to cosine similarity, L2 similarity exhibits a more uniform distribution in the 2D projection, avoiding the clustering concentration caused by the direction sensitivity of cosine similarity. This characteristic ensures more balanced contribution weights across seismic data points, thereby maintaining the balance between clusters.

\subsection{FFN}\label{FFN}
\begin{figure}[!htbp]
  \centering
{\includegraphics[width=0.5\textwidth]{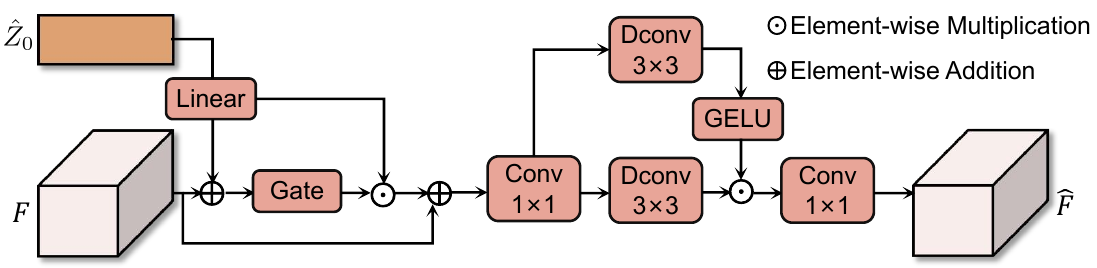}}
\vspace{-5mm}
  \caption{FFN module}
  \label{fig:ffn module}
\vspace{-2mm}
\end{figure}

The SPA module achieves collaborative optimization of local dependencies and global multi-scale features in feature extraction through the structured sparse design and enhanced information extraction. The FFN performs nonlinear enhancement and position-independent processing on the features output by the self-attention mechanism, further enhancing the representational capacity of the features. As Fig. \ref{fig:ffn module} shows, we adopt the same dynamic feature calibration mechanism proposed in Section \ref{sec:Dynamic Feature Calibration Mechanism} to achieve the adaptive fusion of external prior information 
$\hat{\boldsymbol{Z}}_0$, also denoting the fused feature as $\hat{\boldsymbol{F}}_G$. We then use the basic structure of the FFN from the transformer architecture \cite{Xia_2023_ICCV}, which consists of two linear transformations and a nonlinear activation function, and the formula is defined as
\begin{equation}
\hat{\boldsymbol{F}} = \boldsymbol{W}_{f2}\hat{\boldsymbol{F}}_G\odot\operatorname{GELU}
\left(\boldsymbol{W}_{f1} {\hat{\boldsymbol{F}}_G}\right), 
\end{equation}
where $\boldsymbol{W}_{f1}$ and $\boldsymbol{W}_{f2}$ 
represent the two $3\times3$ convolution operations, and $\hat{\boldsymbol{F}}_G$ undergoes $1\times1$ convolutions at both ends to expand and restore the channel dimensions.
It strengthens local detail feature extraction by combining dynamic gating and convolution operations while maintaining the global modeling capability of the Transformer.

This Diff-spaformer framework optimizes the interplay between conditional feature encoding, diffusion-based prior generation, and multi-scale feature integration, ensuring efficient and accurate reconstruction.
Algorithm \ref{alg:Diff-spaformer_training} and Algorithm \ref{alg:Diff-spaformer_interpolation} outline the procedures for the training and inference phases of our model, respectively.
\begin{algorithm}[htbp]
\caption{Training process of Diff-spaformer}
\label{alg:Diff-spaformer_training}
\hspace*{0.02in} {\bf Input:} Training ground truth seismic data $\{\boldsymbol{X}^i\}_{i=1}^n$ with total number $n$; Initializing the SPEN, Spaformer, and the noise-matching network; \\Diffusion steps $T$;
batch size $K$; the number of epochs $N$.
\begin{algorithmic}[1]
\State Randomly initialize Diff-spaformer;
\For{$i=1, \ldots, N$}
    \State Sample batch data $\{\boldsymbol{X}^i\}_{i=1}^K$ from training data;
    \State Generate the binary missing masks $\{\boldsymbol{M}^i\}_{i=1}^K$;
    \State Construct missing data $\{\boldsymbol{X}_{obs}^i\}_{i=1}^K$ according to Eq. \ref{equ:observation_form_complete};
    \State Get the compressed prior feature $\boldsymbol{Z}_0$ form Eq. \ref{equ:spen_x} and  \hspace*{0.16in} the conditional prior information $\boldsymbol{Z}_{c}$ form Eq. \ref{equ:spen_obs};
    \State Create the $T$-steps forward process based on Eq. \ref{equ:q_t_z};
    \For{$t=T, \ldots, 1$}
        \State Get $\{\hat{\boldsymbol{Z}}_{0}^{i,t}\}_{i=1}^K$ from the noise matching process \hspace*{0.36in} according to Eq. \ref{equ:network_eps_pred};
        \State Get $\{\hat{\boldsymbol{Z}}_{t-1}^i\}_{i=1}^K$ from reverse process according to \hspace*{0.36in} Eq. \ref{equ:sampling_z_pred};
    \EndFor
    \State \textbf{end for} 
    \State Get predictions $\{\hat{\boldsymbol{X}}^i\}_{i=1}^K$ from the Spaformer network;
    \State Update the Diff-spaformer network with $\mathcal{L} $ in Eq. \ref{equ:loss_total};\\
\textbf{end for}
\EndFor
\end{algorithmic}
\end{algorithm}

\begin{algorithm}[htbp]
\caption{Inference process of Diff-spaformer}
\label{alg:Diff-spaformer_interpolation}
\hspace*{0.02in} {\bf Input:}
Missing seismic data $\boldsymbol{X}_{obs}$;
Corresponding binary missing mask $\boldsymbol{M}$;
Trained Diff-spaformer model.
\begin{algorithmic}[1]
\State Get the conditional prior information $\boldsymbol{Z}_{c}$ form Eq. \ref{equ:spen_obs};
\State Sample $\boldsymbol{Z}_T$ from the standard normal distribution $\mathcal{N}(\mathbf{0}, \mathbf{I})$;
\State $\hat{\boldsymbol{Z}}_T=\boldsymbol{Z}_T$
\For{$t=T, \ldots, 1$}
    \State Generate $\hat{\boldsymbol{Z}}_{0}^{t}$ from the noise matching process based  \hspace*{0.16in} on Eq. \ref{equ:network_eps_pred};
    \State Get $\hat{\boldsymbol{Z}}_{t-1}$ from reverse process according to Eq. \ref{equ:sampling_z_pred};
    \EndFor
    \State \textbf{end for} 
\State Get predictions $\hat{\boldsymbol{X}}$ from the Spaformer network;
\end{algorithmic}
\hspace*{0.02in} {\textbf{Output:} Interpolated data $\widetilde{\boldsymbol{X}} = \boldsymbol{M}\odot\boldsymbol{X}_{obs}+(1-\boldsymbol{M})\odot\hat{\boldsymbol{X}}$.}
\end{algorithm}

\section{Experiments}\label{secExperiments}

\subsection{Evaluation Metrics}
The fidelity evaluation of interpolated seismic data is quantitatively assessed through three key metrics, including Mean Squared Error (MSE), Signal-to-Noise Ratio (SNR), and Peak Signal-to-Noise Ratio (PSNR). For the interpolated seismic data  $\{\widetilde{\boldsymbol{X}}^i\}_{i=1}^n$ and the ground truth $\{\boldsymbol{X}^i\}_{i=1}^n$, MSE quantifies the average squared deviation as 
\begin{equation}
\operatorname{MSE}=\frac{1}{n}\sum_{i=1}^n({\widetilde{\boldsymbol{X}}^i-\boldsymbol{X}^i})^{2}.
\end{equation}
This metric measures the dispersion of reconstruction errors, with values approaching zero indicating superior reconstruction accuracy and interpolation fidelity. The squared term amplifies significant deviations while preserving dimensional alignment with source data. SNR evaluates the signal preservation capability in decibels (dB) through power ratio analysis as
\begin{equation}
\operatorname{SNR}=10\log_{10}\frac{\|\boldsymbol{X}\|^{2}_{F}}{\|\boldsymbol{X}-\widetilde{\boldsymbol{X}}\|^{2}_{F}},
\end{equation}
where $\|\cdot\|_F$ refers to the Frobenius norm. PSNR extends SNR by incorporating the dynamic range of seismic signals and is defined as
\begin{equation}
\operatorname{PSNR}=10\log_{10}\frac{\operatorname{MAX}_{\boldsymbol{X}}^2}{\operatorname{MSE}},
\end{equation}
where $\operatorname{MAX}_{\boldsymbol{X}}$ represents the maximum amplitude of $\boldsymbol{x}_\text{gt}$. Higher SNR and PSNR values correspond to superior interpolation quality.
We use the Structural Similarity Index (SSIM)  \cite{zhou1284395} to evaluate the structural similarity between two seismic datasets by comparing their local statistical features. The calculation formula of SSIM is as follows 
\begin{equation*}
\operatorname{SSIM}(\boldsymbol{X}, \widetilde{\boldsymbol{X}})=\frac{\left(2 \mu_{\boldsymbol{X}} \mu_{\widetilde{\boldsymbol{X}}}+C_1\right)\left(2 \sigma_{\boldsymbol{X} \widetilde{\boldsymbol{X}}}+C_2\right)}{\left(\mu_{\boldsymbol{X}}^2+\mu_{\widetilde{\boldsymbol{X}}}^2+C_1\right)\left(\sigma_{\boldsymbol{X}}^2+\sigma_{\widetilde{\boldsymbol{X}}}^2+C_2\right)},
\end{equation*}
where $\mu_{\boldsymbol{X}} (\mu_{\widetilde{\boldsymbol{X}}})$, $\sigma_{\boldsymbol{X}}(\sigma_{\widetilde{\boldsymbol{X}}})$, and $2 \sigma_{\boldsymbol{X} \widetilde{\boldsymbol{X}}}$ denote the mean (luminance), variance (contrast), and covariance (structure), respectively. $C_1$ and $C_2$ are constants introduced to avoid division by zero, typically set to very small values (e.g., $C_1=1e-4$, $C_2=1e-4$).
In practical applications, SSIM is computed by sliding a window (e.g., a $3\times3$ pixel window) across the seismic data, calculating the value for each block, and then averaging the results over the entire data to capture local spatial characteristics. The resulting similarity value ranges from -1 to 1, with values closer to 1 indicating greater structural similarity.
This multi-metric framework enables comprehensive characterization of reconstruction accuracy, noise suppression effectiveness, amplitude preservation fidelity, and structural similarity in seismic data interpolation tasks.

\begin{table*}[htbp]
\renewcommand{\arraystretch}{1.2} 
\setlength{\tabcolsep}{6pt} 
\caption{Comparison of various methods on the test set of different datasets with random missing traces. The best results are highlighted in bold.}
\centering
\begin{tabular}{lrrrrrrrrrrrrrrrr}
\toprule 
\textbf{Dataset} & \multicolumn{4}{c}{SEG C3} & \multicolumn{4}{c}{MAVO} & \multicolumn{4}{c}{Model94}\\
Model & MSE$\downarrow$ & SNR$\uparrow$ & PSNR$\uparrow$ &SSIM$\uparrow$ & MSE$\downarrow$ & SNR$\uparrow$ & PSNR$\uparrow$ &SSIM$\uparrow$& MSE$\downarrow$ & SNR$\uparrow$ & PSNR$\uparrow$  &SSIM$\uparrow$\\
\midrule  
\textbf{DD-CGAN}\cite{Chang2020}  & 2.284e-04 & 30.458 & 36.412 & 0.952 & 2.639e-04 & 30.254 & 35.786 & 0.955 & 2.421e-04 & 29.588 & 36.160 & 0.978\\ 
\textbf{cWGAN-GP}\cite{wei2022big}   & 9.708e-05 & 34.175 & 40.129 & 0.981 & 1.971e-04 & 31.522 & 37.054 & 0.967 & 1.390e-04 & 31.997 & 38.569 & 0.987\\
\textbf{PConv-UNet}\cite{pan2020partial}  & 6.810e-05 & 35.715 & 41.669 & 0.986 & 1.221e-04 & 33.601 & 39.133  & 0.977 & 1.870e-04 & 30.711 & 37.283 & 0.984\\ 
\textbf{ANet}\cite{Yu9390348} & 1.204e-04 & 33.240 & 39.194 & 0.977 & 1.889e-04 & 31.705 & 37.237 & 0.969 & 1.562e-04 & 31.493 & 38.065 & 0.986\\
\textbf{Coarse-to-Fine}\cite{wei2022hybrid} & 7.260e-05 & 35.437 & 41.391 & 0.985 & 1.595e-04 & 32.440 & 37.972 & 0.971 & 1.187e-04 & 32.684 & 39.256 & 0.988\\ 
\textbf{SeisFusion*}\cite{10681481} & 9.603e-05 & 34.222 & 40.176 & 0.980 & 1.841e-04 & 31.817 & 37.349 & 0.965 & 1.650e-04 & 31.252 & 37.824 & 0.980\\ 
\textbf{SeisDDIMCR} \cite{10731876}& 4.777e-05 & 37.255 & 43.209  & 0.989 &1.127e-04 & 33.950 & 39.482 & 0.976 & 8.043e-05 & 34.374 & 40.946 & 0.991 \\  
\textbf{Ours} & \bf{3.763e-05} & \bf{38.291} & \bf{44.245} & \bf{0.992} & \bf{1.027e-04} & \bf{34.353} & \bf{39.885} & \bf{0.978} & \bf{4.440e-05} & \bf{36.954} & \bf{43.526} & \bf{0.996}\\ 
\bottomrule
\end{tabular}
  \label{tab:randommissing}
  \vspace{-2mm}
\end{table*}

\subsection{Data Set}
We validate the proposed model on three publicly available datasets, including the synthetic dataset Model94 and Society of Exploration Geophysicists (SEG) C3, and the field dataset Mobil Avo Viking Graben Line 12 (MAVO), all of which are commonly used for seismic data reconstruction tasks. 

The Model94 dataset contains 277 shot-gathers, of which 198 consecutive and complete gathers (each comprising 480 traces with 15 m intervals) are selected following the operation in \cite{song2023regeneration}. Each trace comprises 2,000 time samples at a 4 ms sampling rate. To mitigate partial temporal vacancies, we crop the temporal dimension to 1,000 samples. The data are randomly divided into 118, 40, and 40 shot-gathers for training, validation, and testing, respectively. Due to limited data volume, the dataset is augmented by duplication, yielding 30,000 training, 6,000 validation, and 6,000 test patches through randomized cropping on shot-gathers.

The SEG C3 dataset includes 45 shots, each with a 201 × 201 receiver grid (dx, dy = 20 m) and 625 samples per trace (dt = 8 ms). Due to the limited valid seismic events in the bottom part of time samples, we select parts of the data for our experiment, resulting in a final data size of 45$\times$201$\times$201$\times$300. Following the training, validation, and test set generation rules used in \cite{10731876}, for each seismic shot, we randomly selected 20 gathers for validation and 20 gathers for testing along the inline direction, with the remaining 161 gathers used for training. Then, on each slice, patches with both the time and trace dimensions of size 128 are randomly cropped. Ultimately, the total numbers of patches for validation, testing, and training are 30,000, 6,000, and 6,000, respectively.

The MAVO dataset consists of a 1001$\times$120 receiver grid and 1500 time samples per trace, recorded with a time interval of 4 ms and a trace interval of 25 m. We exclude some of the later received data and construct 1001$\times$120$\times$1000 data for experiments. Following \cite{10731876}, we randomly divide all the gathers into three parts, i.e., 801 for training, 100 for validation, and the remaining 100 for testing. Then, patches are randomly cropped from each part, resulting in 20,000 training patches, 4,000 validation patches, and 4,000 testing patches. Each patch keeps time and trace dimensions of 256 and 112, respectively.

Before being fed into models, all seismic patches are first normalized to the range $[0,1]$ using min-max normalization.

\subsection{Comparison Method}
We compare our model with 7 methods, including two GAN-based approaches, i.e., the dual-domain optimized DD-CGAN \cite{Chang2020} and conditional Wasserstein generative adversarial networks with gradient penalty (cWGAN-GP) \cite{wei2022big}, partial convolution-based U-Net (PConv-UNet) \cite{pan2020partial}, CNN guided by attention
mechanism (ANet) \cite{Yu9390348}, two-stage Coarse-to-Fine model \cite{wei2022hybrid}, and diffusion-based models SeisFusion \cite{10681481} and SeisDDIMCR \cite{10731876}. SeisDDIMCR and SeisFusion share the same backbone network. SeisFusion* (differentiated by the asterisk) implements single-network noise matching to eliminate dual-network computational costs. This critical divergence from the original dual-network architecture is explicitly marked in the notation. All other hyperparameter configurations follow their respective architectures.

\subsection{Implementation Details}
In the diffusion model configuration, the number of diffusion steps is set to $4$, and the noise variance coefficient $\beta$ increases linearly from $0.1$ to $0.99$. The Spaformer has a four-layer channel dimension, increasing from $64$, $64\times2$, $64\times4$, to $64\times8$. The dimension 
$C^{\prime}$ of the compressed prior information is set to $64\times4$. The weight coefficients of the loss function are set to $\lambda_1=6.0$, $\lambda_2=1.0$, and $\lambda_3=6.0$. We execute Algorithm \ref{alg:Diff-spaformer_training} using the Adam (Adaptive Moment Estimation) gradient descent optimization algorithm to train the model on different datasets. We train every model separately for two missing data scenarios, including random and continuous missing types. The random missing rate is set between 0.2 and 0.8, and the continuous missing rate ranges from 0.1 to 0.6. To preserve edge integrity, we prevent continuous missing cases from occurring near boundary traces. Masks are randomly generated during each training iteration to ensure sufficient diversity in training sample pairs. The training consists of $100$ epochs with a batch size of $20$. The learning rate follows a piecewise constant decay strategy, starting at $1e-4$, and decreasing by a factor of $10$ after $50$ epochs. The model parameters for the comparison methods are set to be consistent with those in the original paper, while the training strategy is aligned with that of our model. However, the total number of training epochs differs between the comparison methods and our approach, with the comparison methods undergoing more training epochs than our method. We conduct total experiments on PyTorch 1.12.1 and NVIDIA A100 Tensor Core GPU.

\subsection{Experimental Results}\label{section:Experimental Results}
Random missing traces and continuous missing traces are two common manifestations of seismic data gaps. High-density random missing traces can lead to data sparsity, introducing aliasing artifacts and spectral leakage, and models need to implicitly learn anti-aliasing capabilities. Continuous missing traces rely more on global semantic reasoning, and it is crucial to reduce distribution discrepancies between missing and non-missing regions.  We implement Algorithm \ref{alg:Diff-spaformer_interpolation} to validate the performance of our model. Similarly, the comparative methods are also tested on the same dataset.

\begin{figure*}[htbp]
  \centering
    \subfloat[Ground Truth.]
  {\includegraphics[height=0.19\textwidth]{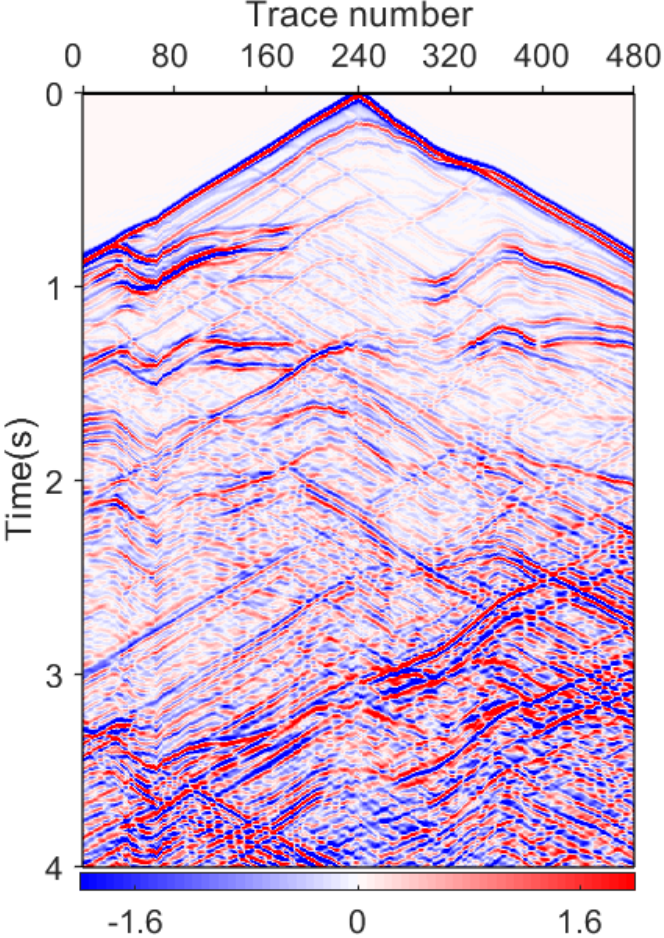}\label{fig:model94random_subfig1}\hspace{1.0mm}}
  \subfloat[DD-CGAN.]
  {\includegraphics[height=0.19\textwidth]{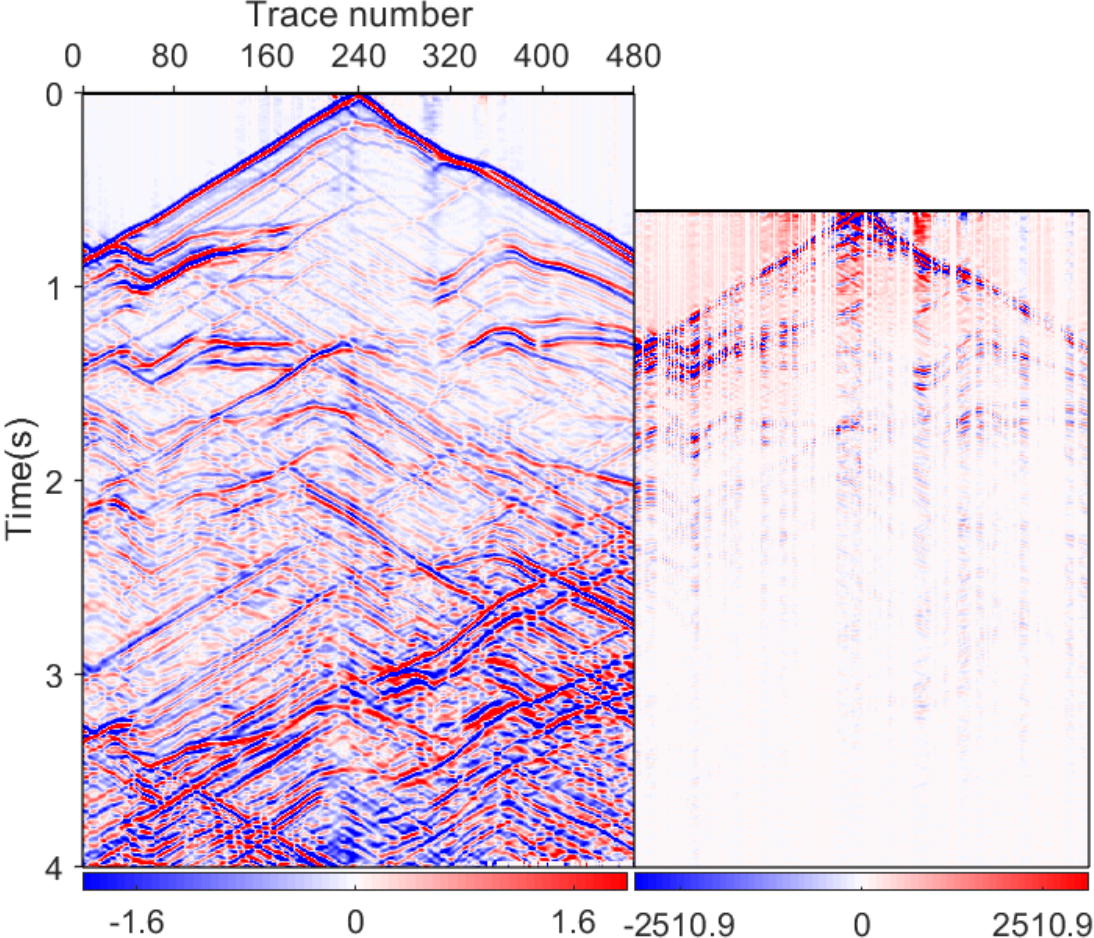}\label{fig:model94random_subfig3}}
  \subfloat[cWGAN-GP.]
  {\includegraphics[height=0.19\textwidth]{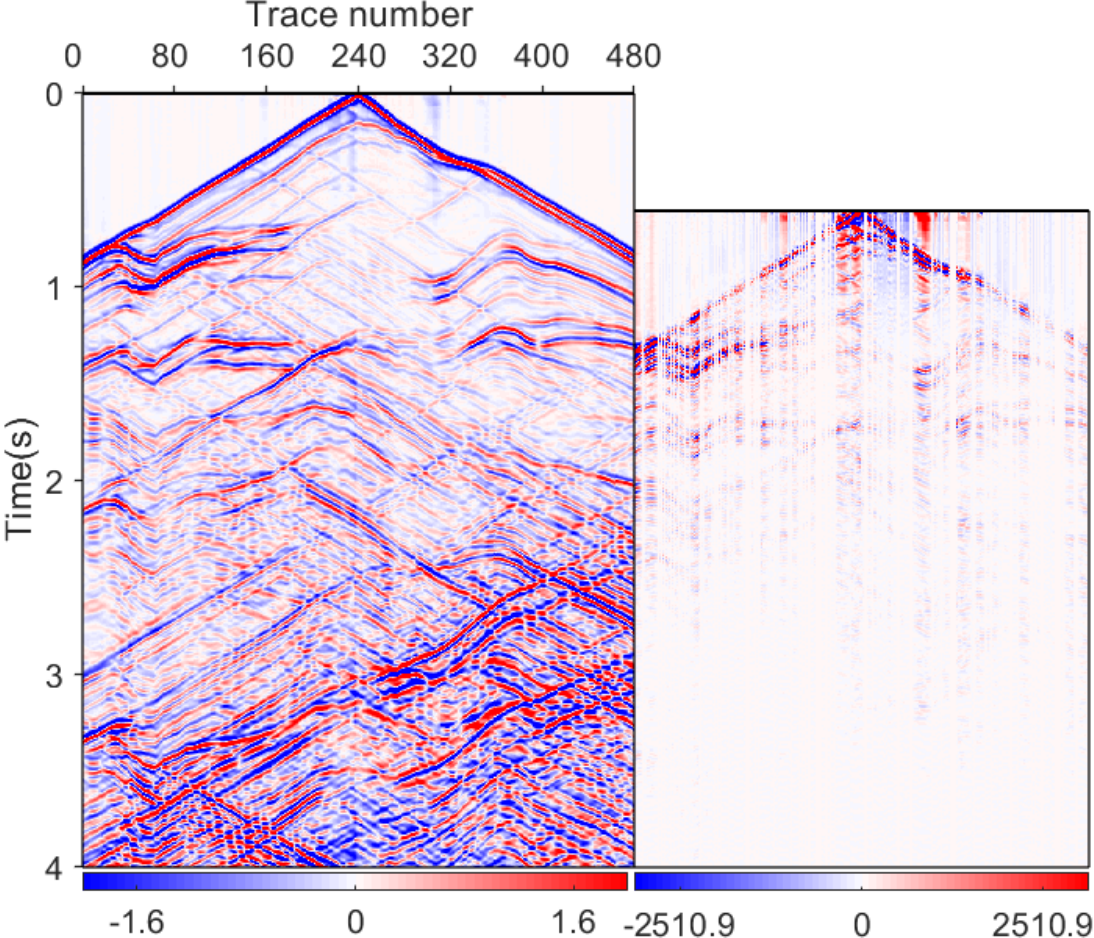}\label{fig:model94random_subfig4}}
  \subfloat[PConv-UNet.]
  {\includegraphics[height=0.19\textwidth]{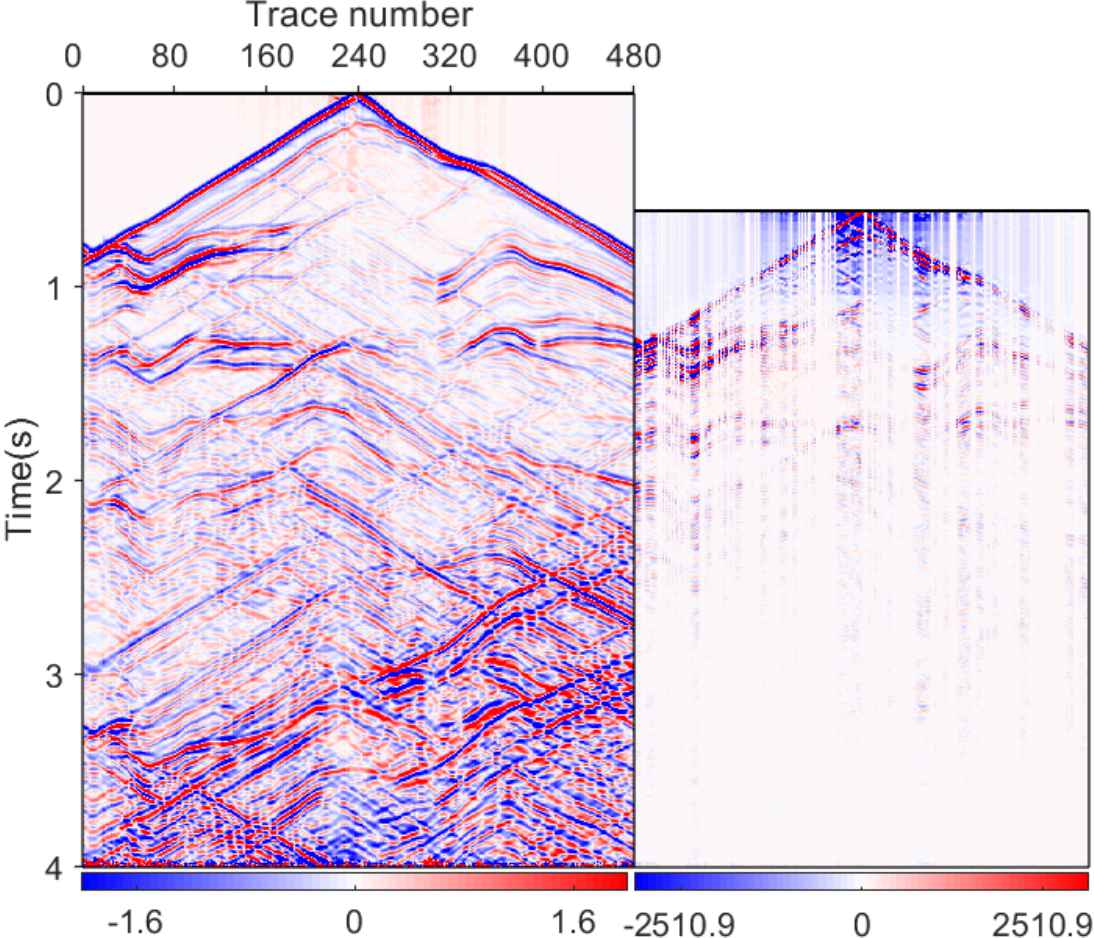}\label{fig:model94random_subfig5}}
  \subfloat[ANet.]
  {\includegraphics[height=0.19\textwidth]{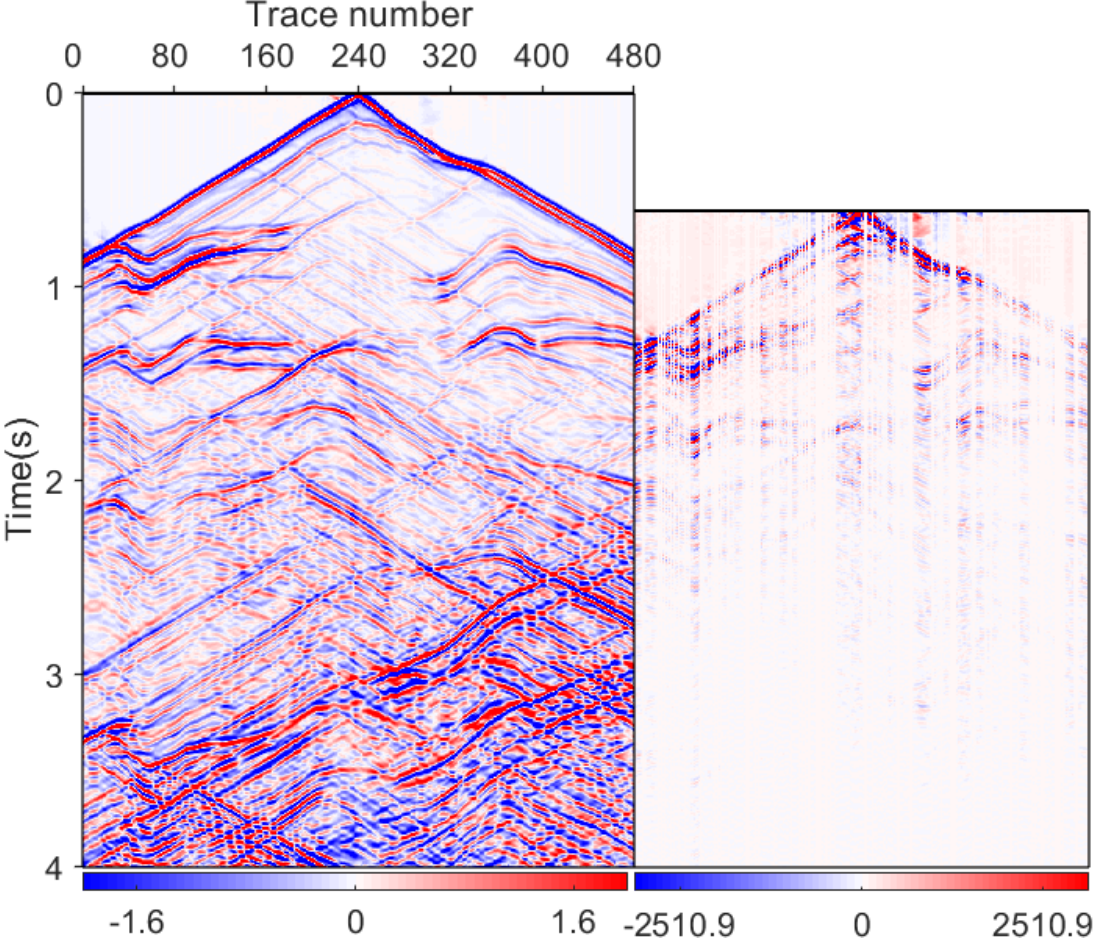}\label{fig:model94random_subfig6}}
 \clearpage
  \subfloat[Missing data.]
{\includegraphics[height=0.19\textwidth]{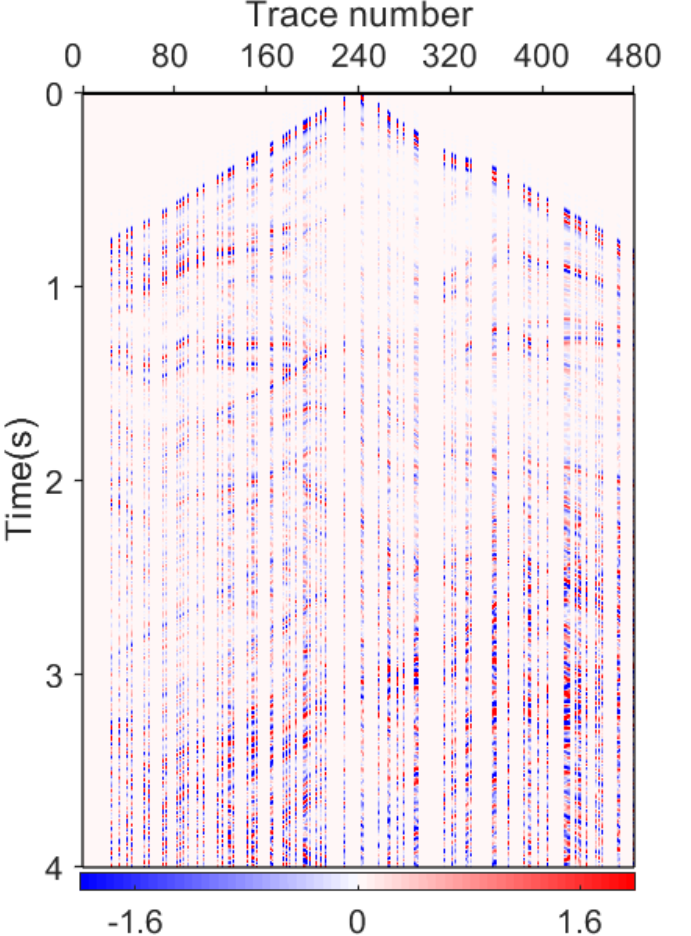}\label{fig:model94random_subfig2}}
  \subfloat[Coarse-to-Fine.]  {\includegraphics[height=0.19\textwidth]{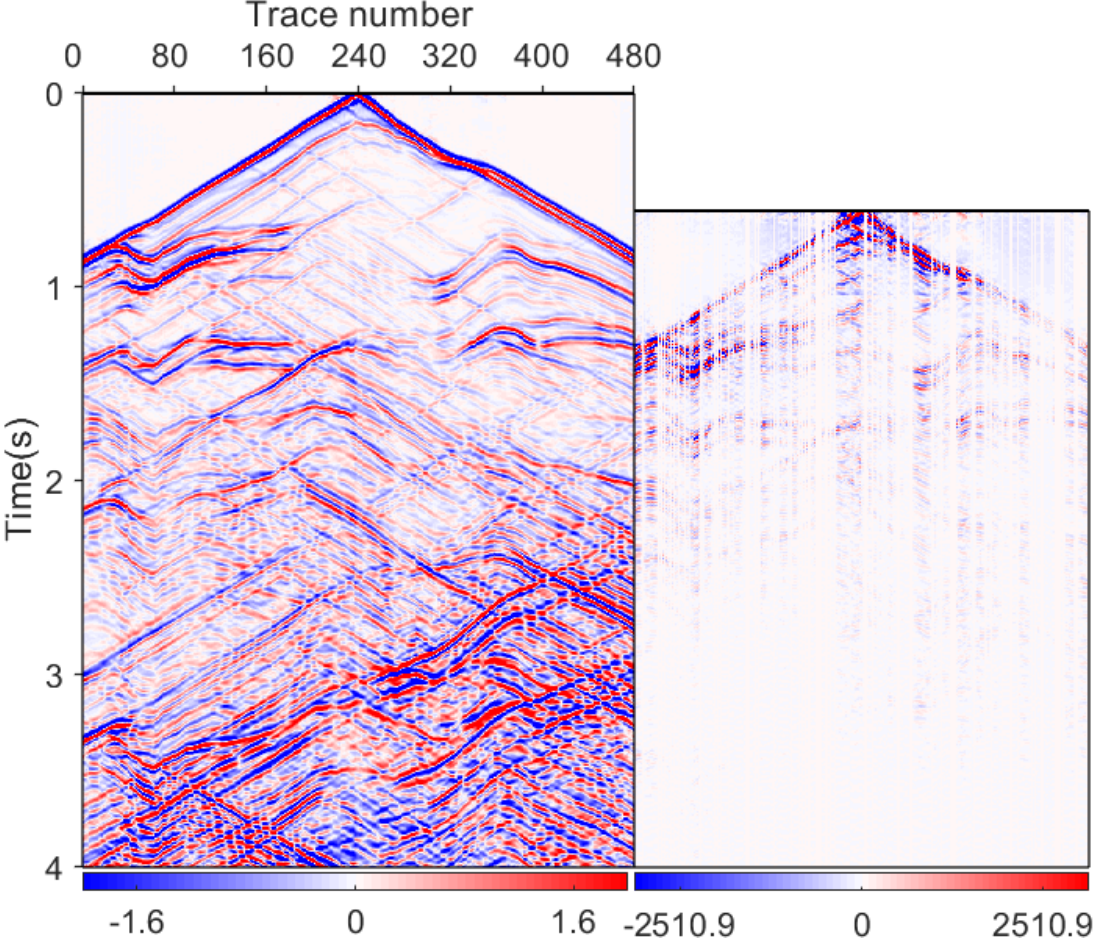}\label{fig:model94random_subfig7}}
  \subfloat[SeisFusion*.]
  {\includegraphics[height=0.19\textwidth]{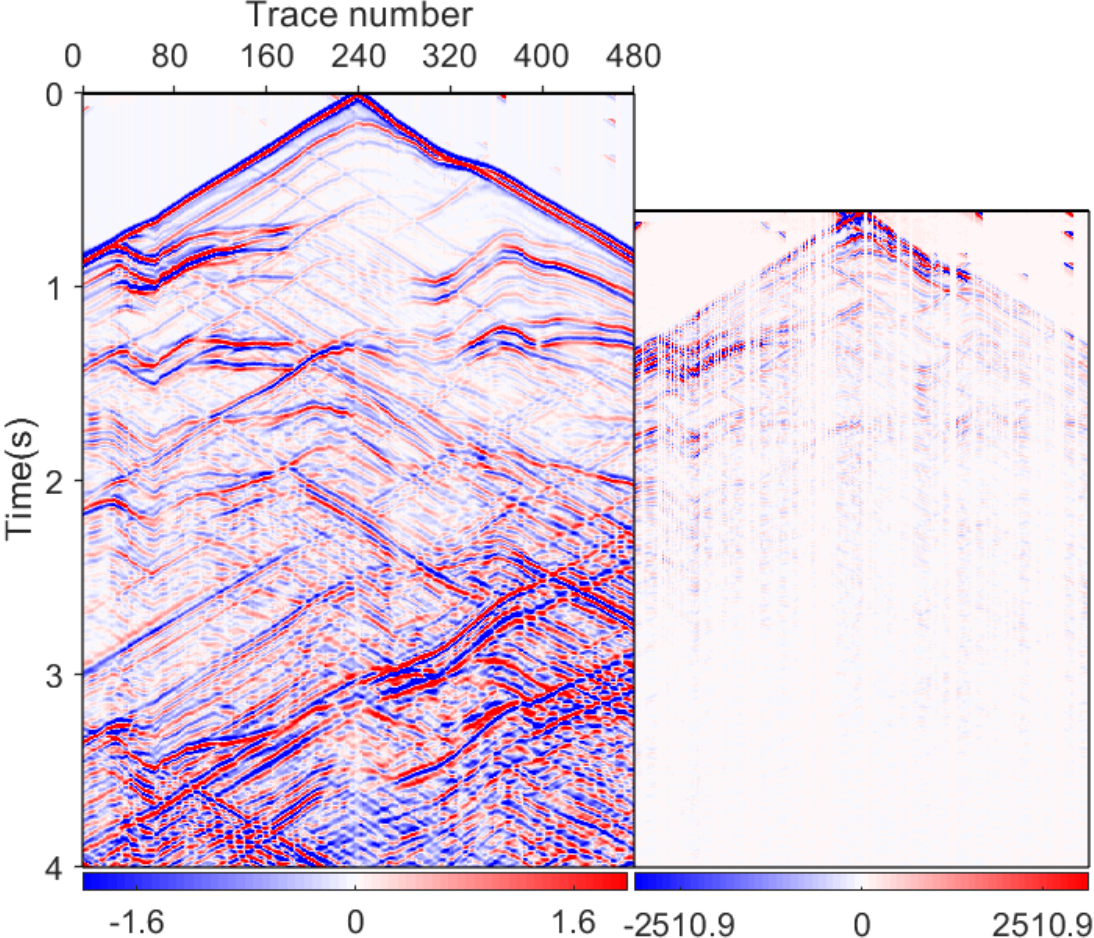}\label{fig:model94random_subfig8}}
  \subfloat[SeisDDIMCR.] 
    {\includegraphics[height=0.19\textwidth]{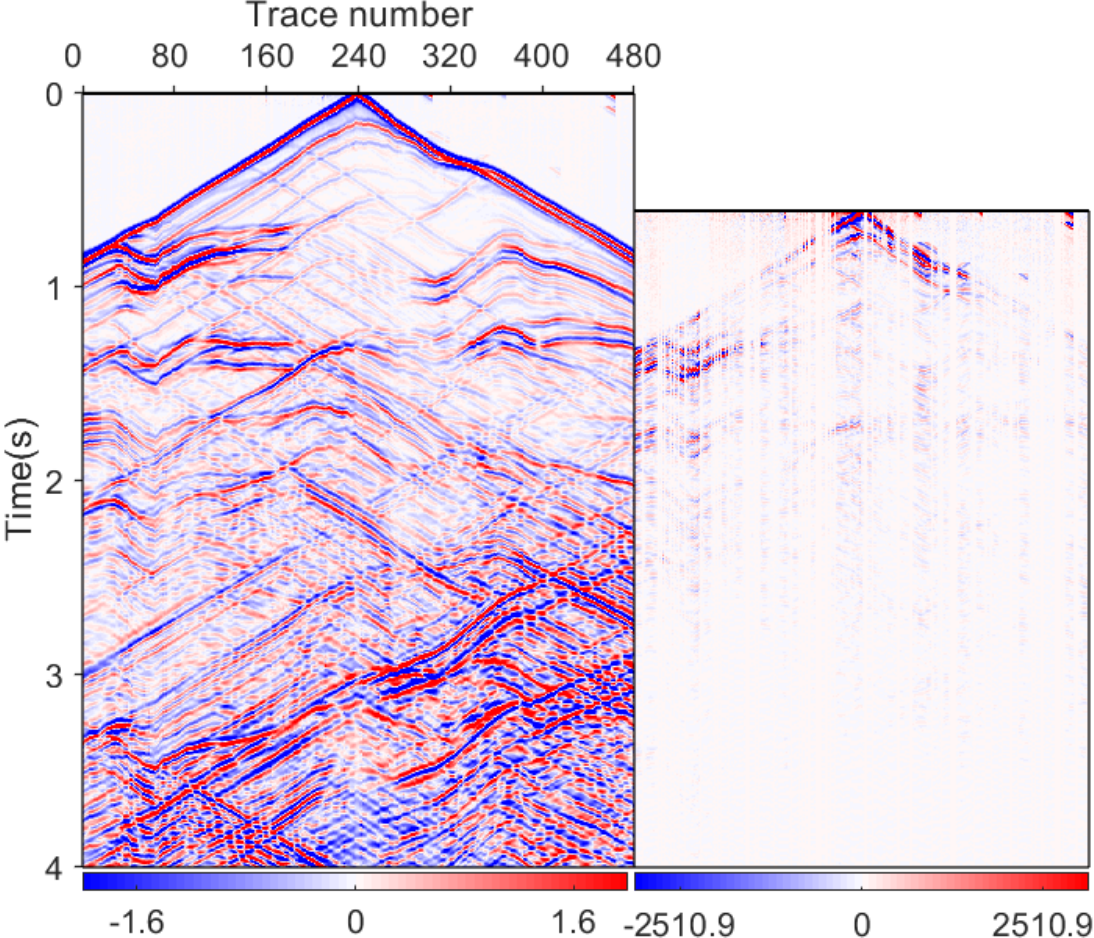}\label{fig:model94random_subfig9}}
  \subfloat[Ours.]
    {\includegraphics[height=0.19\textwidth]{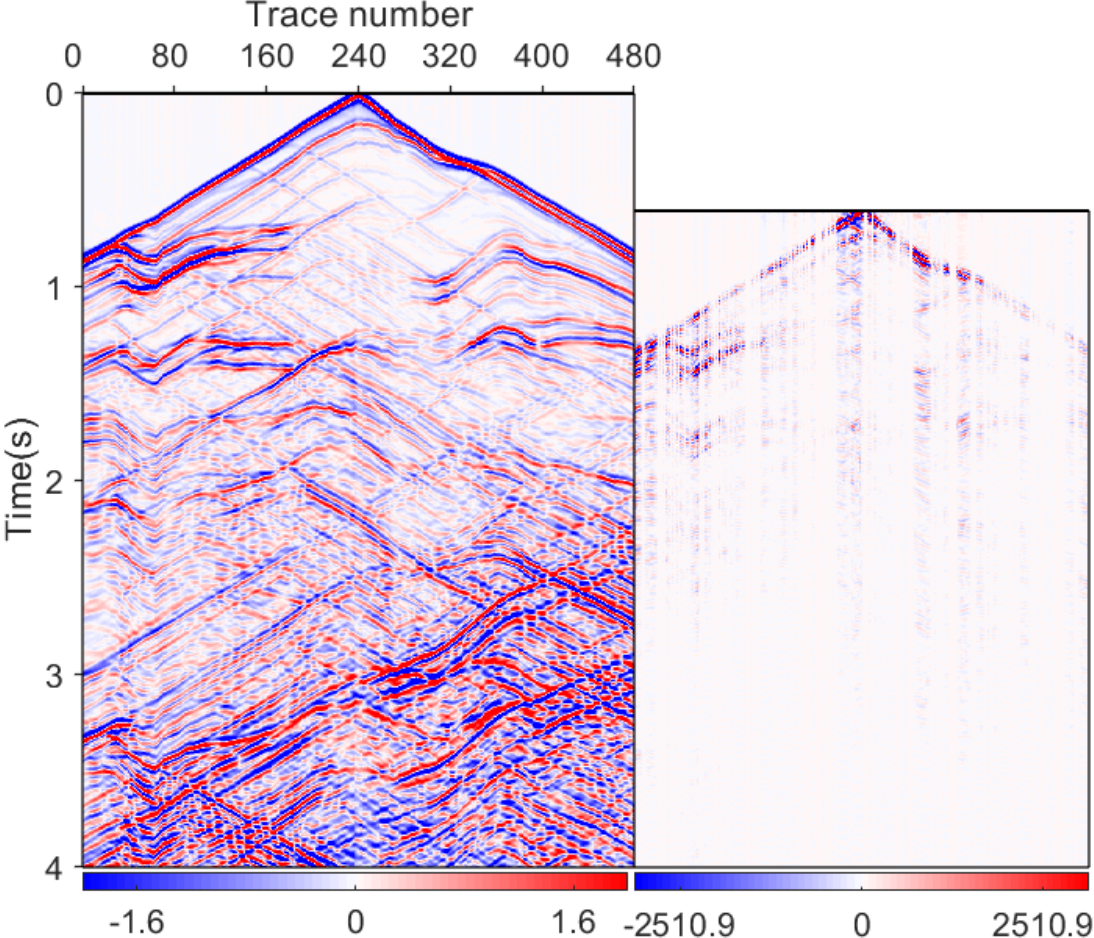}\label{fig:model94random_subfig10}}
  
  \caption{Interpolation results of the Model94 complete test slice with a 79.2\% random missing ratio on different methods. We restore the seismic data to its original amplitude range and apply the gain method to enhance the visibility of weak amplitude details. The reconstruction residuals are displayed on the right panel for better comparison. Residuals are calculated and presented directly from raw data without any gain processing.}
  \label{fig:Model94 random}
\vspace{-4mm}
\end{figure*}
 
\begin{figure*}[htbp]
    \subfloat[Ground Truth.]
  {\includegraphics[height=0.145\textwidth]{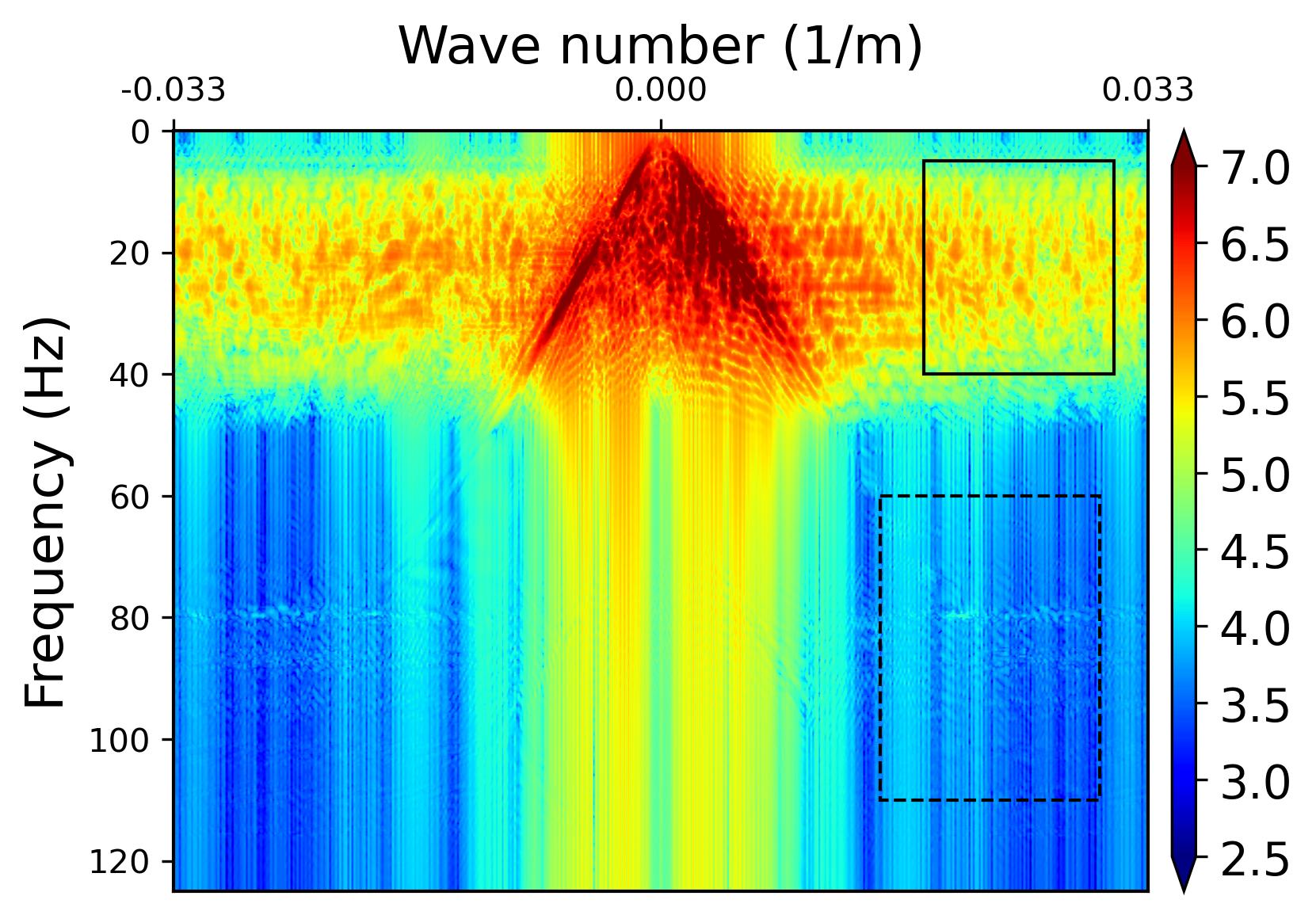}\label{fig:model94randomfk_subfig1}\hspace{1.0mm}}
  \subfloat[DD-CGAN.]
  {\includegraphics[height=0.145\textwidth]{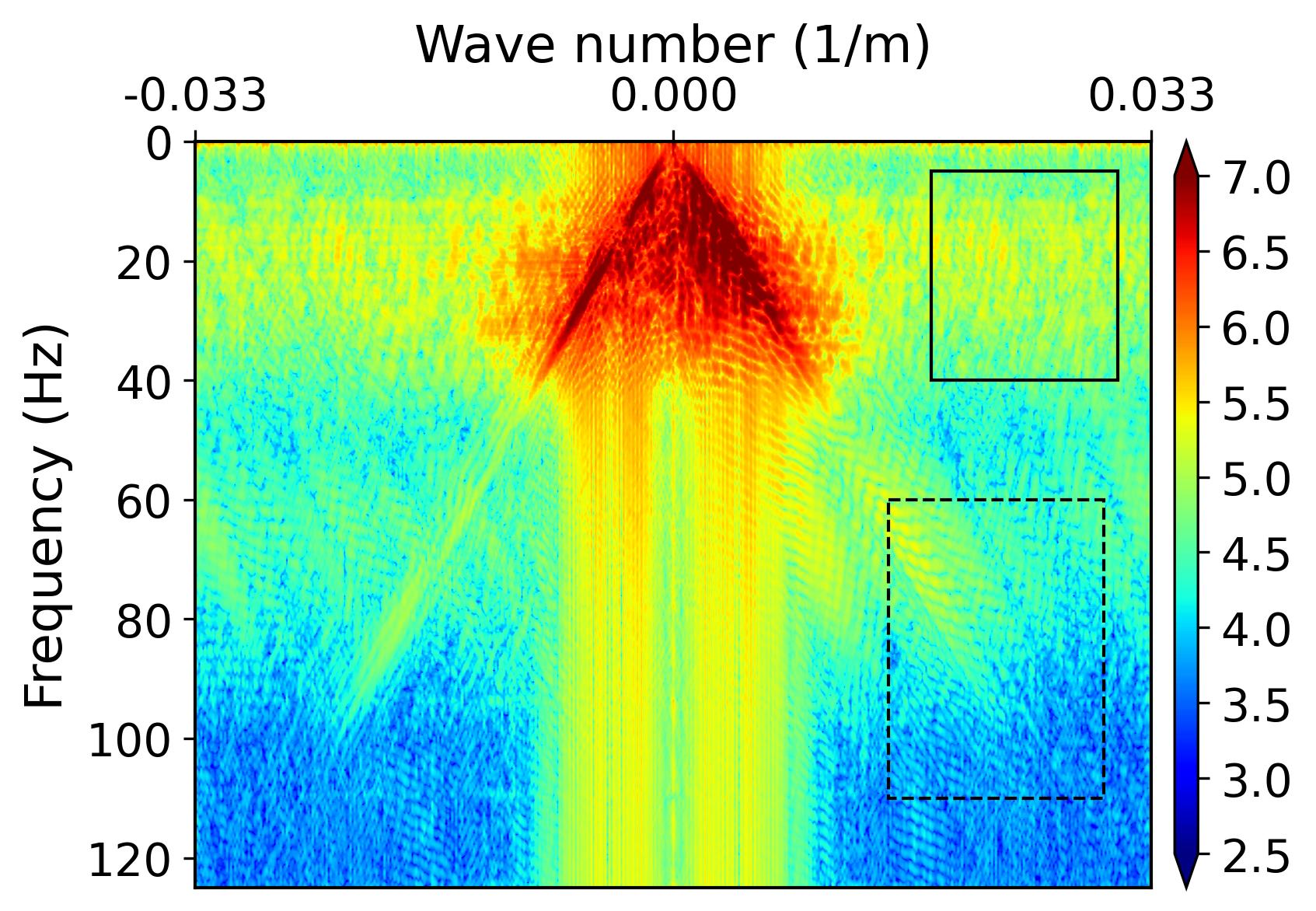}\label{fig:model94randomfk_subfig3}}
  \subfloat[cWGAN-GP.]
  {\includegraphics[height=0.145\textwidth]{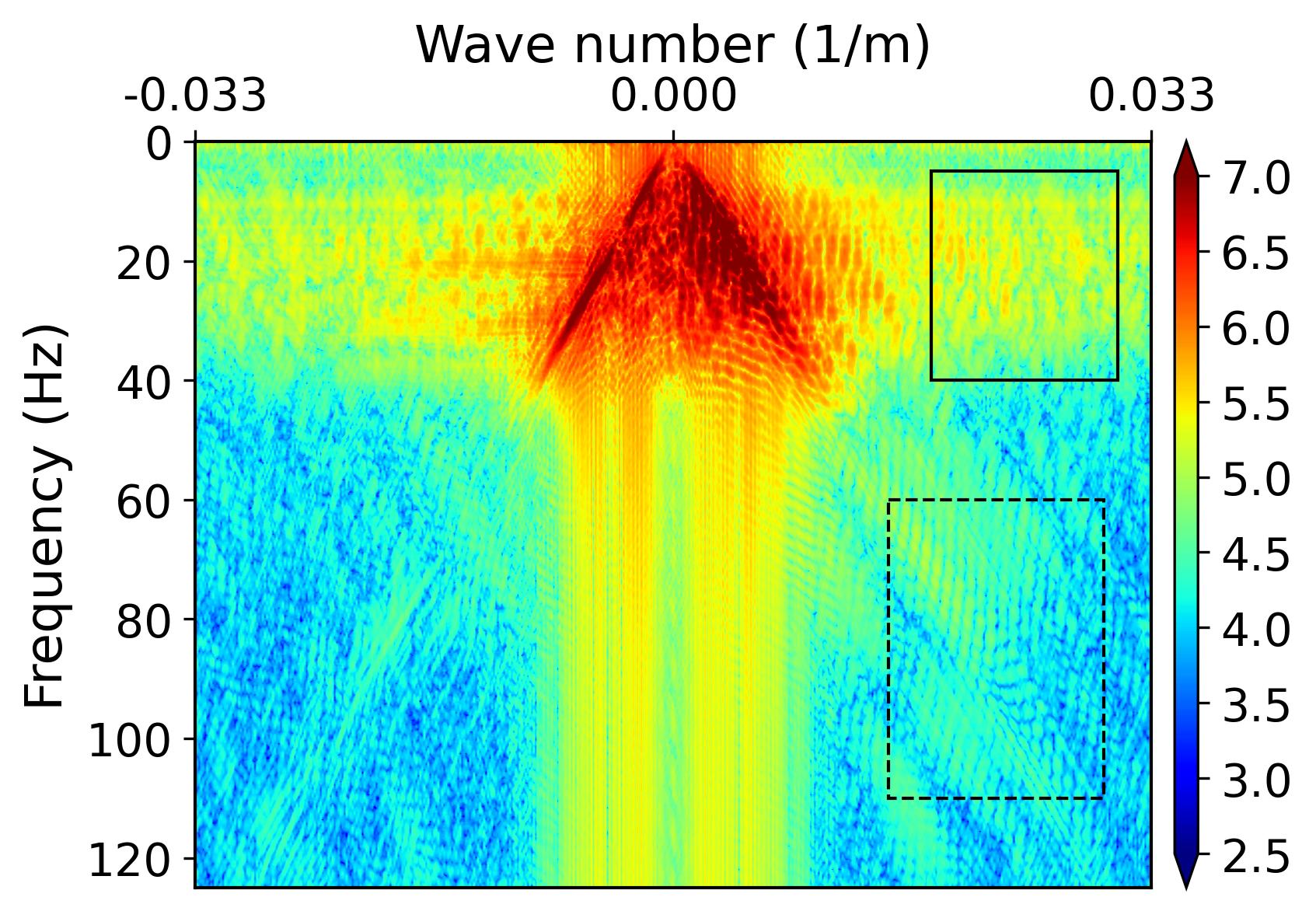}\label{fig:model94randomfk_subfig4}}
  \subfloat[PConv-UNet.]
  {\includegraphics[height=0.145\textwidth]{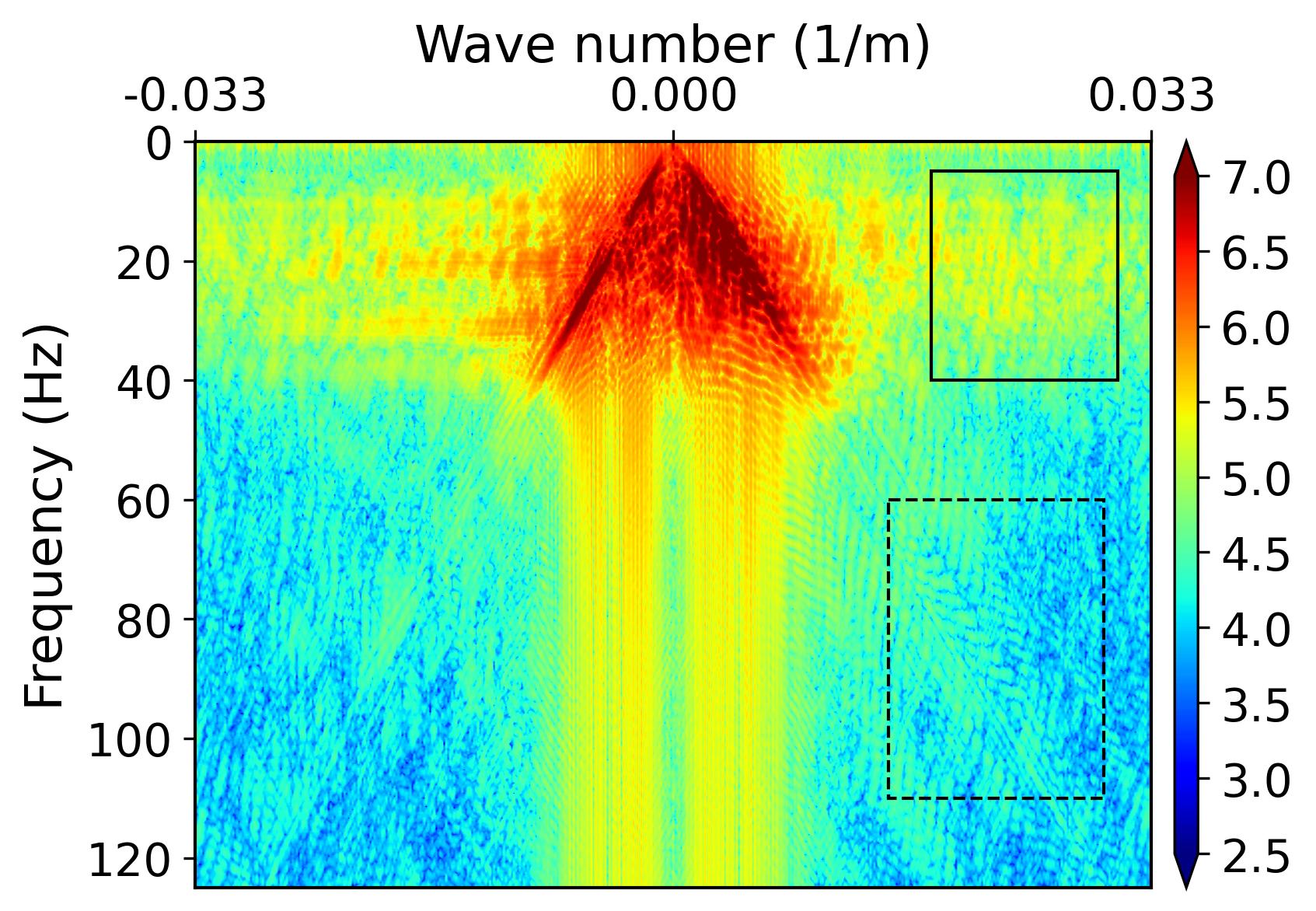}\label{fig:model94randomfk_subfig5}}
  \subfloat[ANet.]
  {\includegraphics[height=0.145\textwidth]{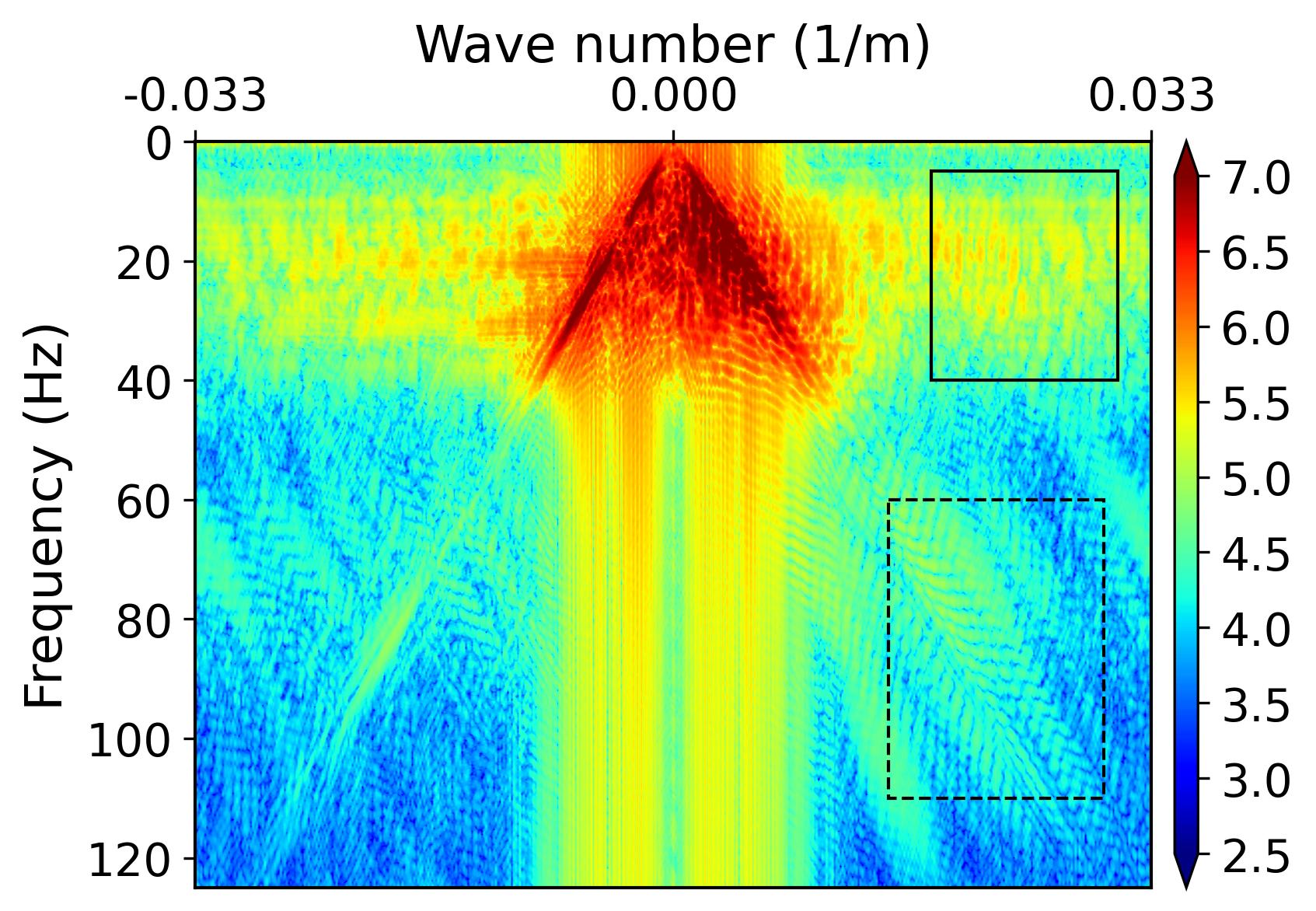}\label{fig:model94randomfk_subfig6}}
 \clearpage
  \subfloat[Missing data.]
{\includegraphics[height=0.145\textwidth]{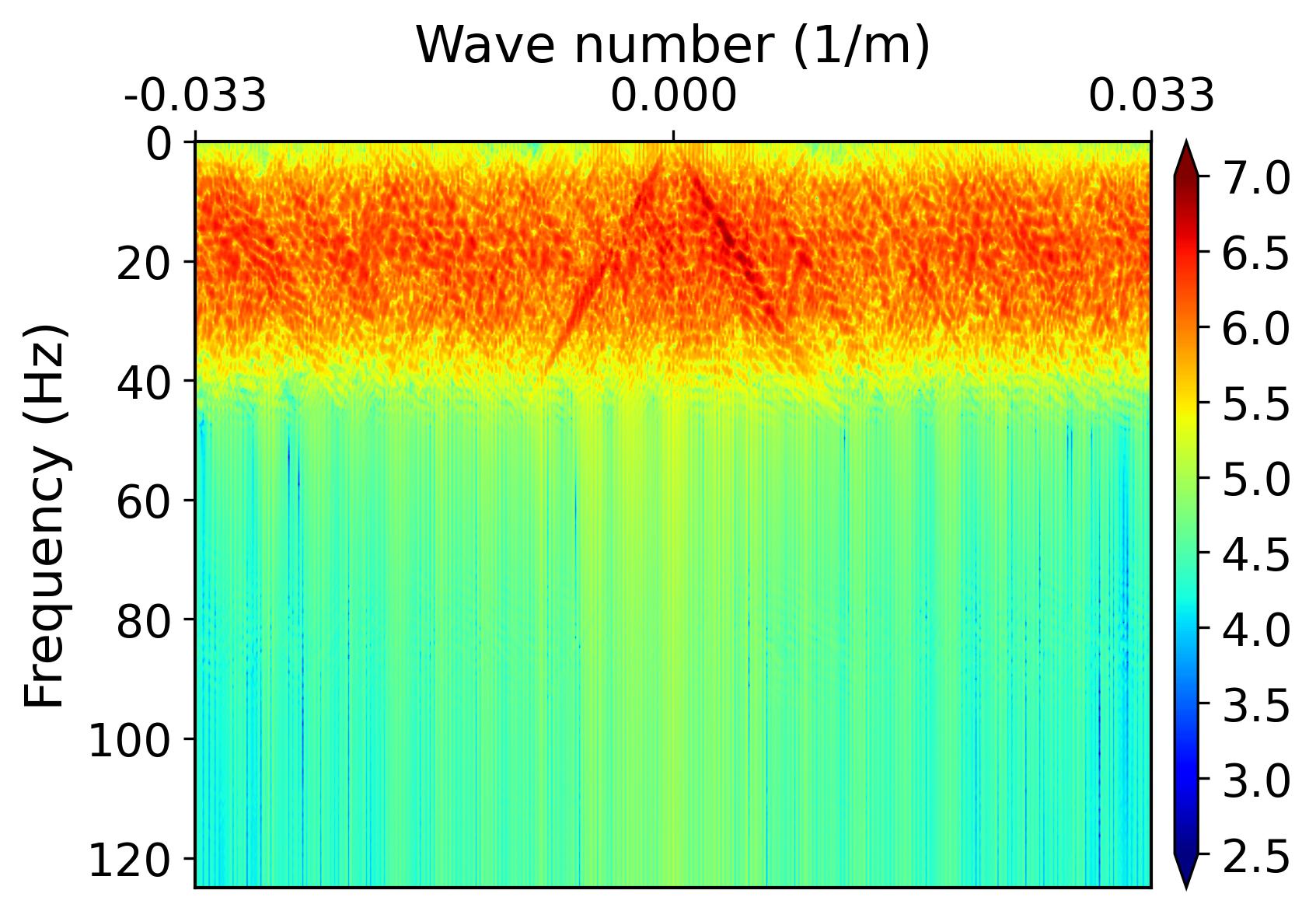}\label{fig:model94randomfk_subfig2}}\hspace{0.3mm}
  \subfloat[Coarse-to-Fine.]  {\includegraphics[height=0.145\textwidth]{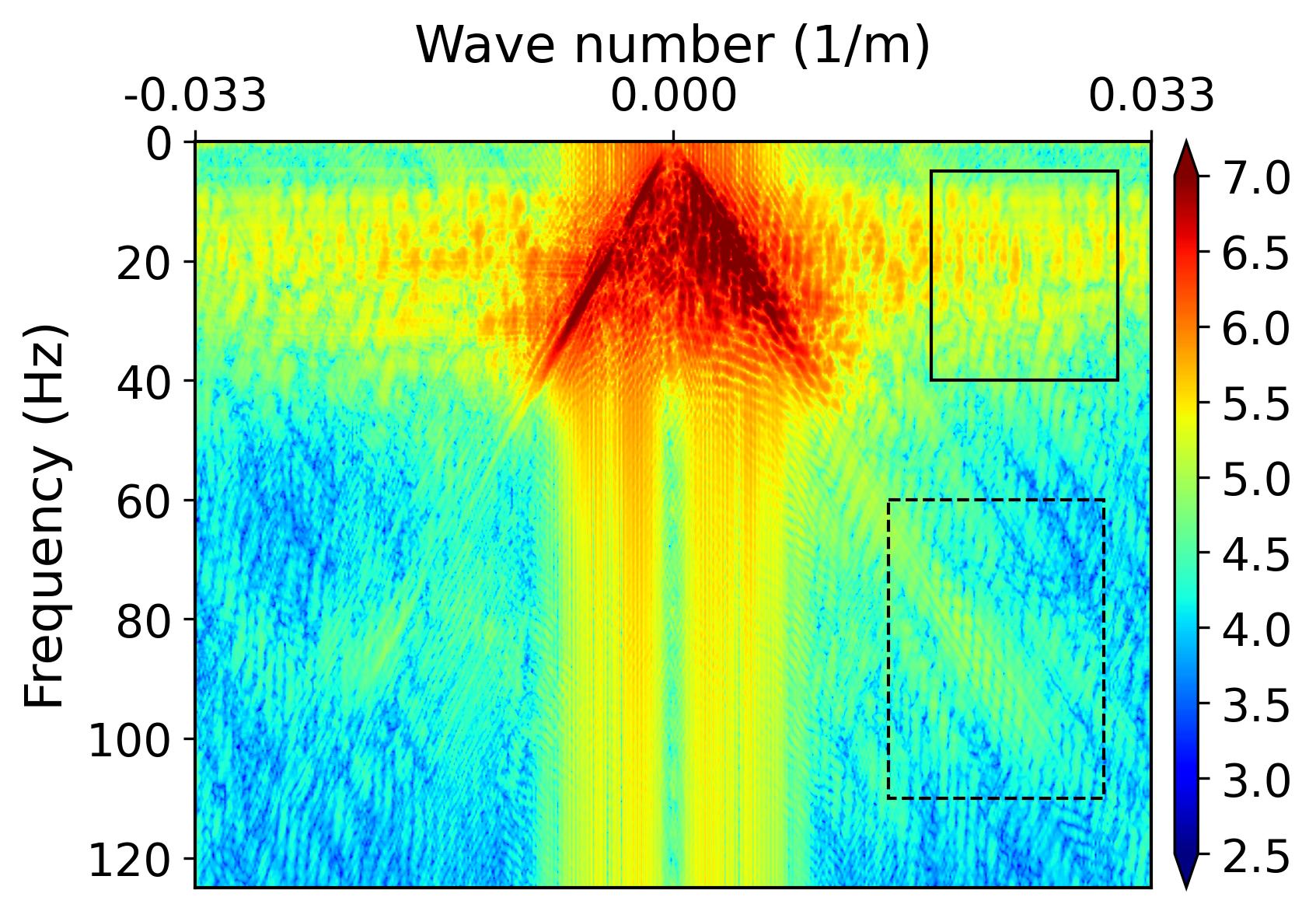}\label{fig:model94randomfk_subfig7}}
  \subfloat[SeisFusion*.]
  {\includegraphics[height=0.145\textwidth]{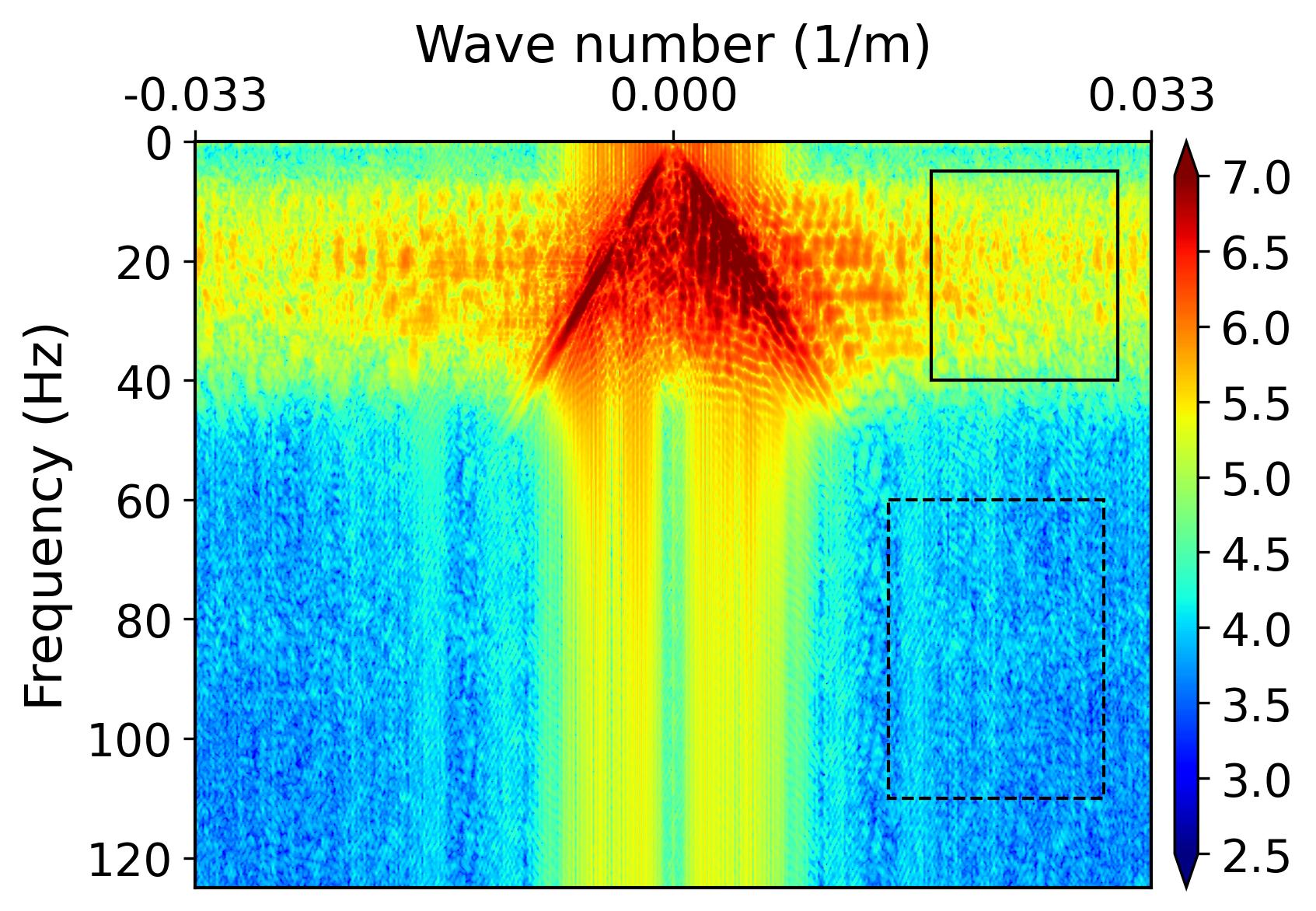}\label{fig:model94randomfk_subfig8}}
  \subfloat[SeisDDIMCR.] 
    {\includegraphics[height=0.145\textwidth]{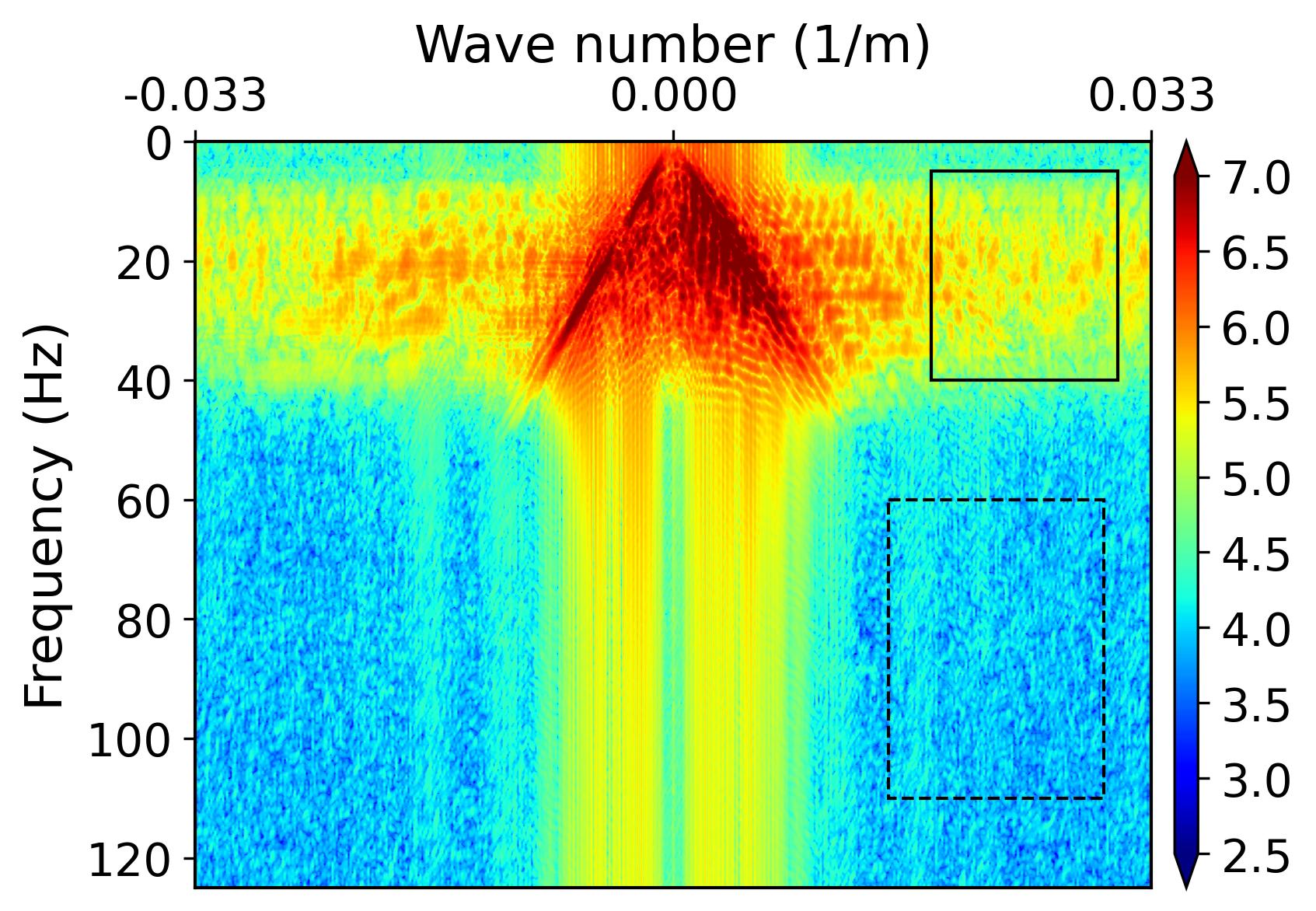}\label{fig:model94randomfk_subfig9}}
  \subfloat[Ours.]
    {\includegraphics[height=0.145\textwidth]{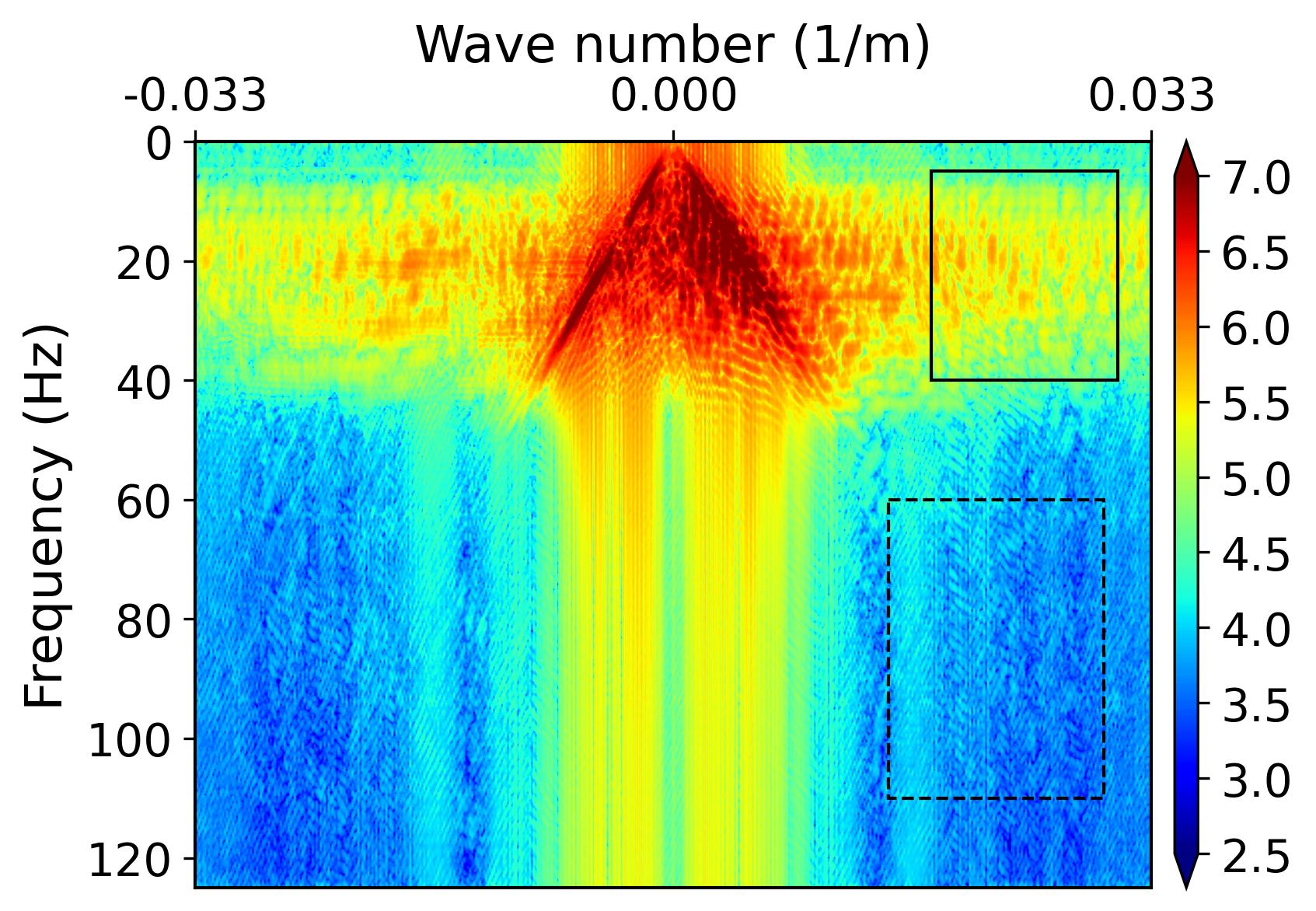}\label{fig:model94randomfk_subfig10}}
  
  \caption{The $f\text{-}k$ spectra of Model94 test data interpolation results with 79.2\% random missing traces on different methods. }
  \label{fig:Model94random fk}
\vspace{-4mm}
\end{figure*}

\subsubsection{Random Missing Traces}\label{exp:Random Missing Traces}
The test random missing rate is consistent with the training set, also ranging from 0.2 to 0.8. Tab. \ref{tab:randommissing} summarizes the quantitative interpolation evaluation results on the three test sets. Compared with comparative methods, our method consistently demonstrates superior performance across all evaluation metrics. It should be noted that, compared to SeisDDIMCR, our method achieves a more than 1 dB improvement in SNR on the SEG C3 and Model94 test sets, demonstrating its ability to bypass the highly iterative interpolation process while maintaining reconstruction fidelity. Fig. \ref{fig:Model94 random} visually compares the interpolation results and residuals of highly sparse random missing traces (missing rate: 79.2\%) on the Model94 complete test slice. Reconstructing complete data from only 20.8\% of the original observations requires joint modeling of global and local correlations while balancing anti-aliasing constraints and high-frequency reconstruction. First, except for Coarse-to-Fine and our proposed method, all other methods exhibit some degree of artifact residue in high-amplitude regions. Second, the residual plots clearly reflect the amplitude errors between the reconstructed signals and the ground truth signals at the same scale. Compared to the other six methods, SeisDDIMCR and our method show noticeably cleaner residual plots. Furthermore, our method exhibits less residual in the middle of the spread (e.g., from trace number 160 to 320). Besides, to quantitatively compare the effectiveness of different methods in suppressing aliasing artifacts induced by highly sparse data, we visualize the corresponding $f\text{-}k$ spectra in Fig. \ref{fig:Model94random fk}. We clearly observe spectral leakage and coherent noise contamination in the $f\text{-}k$ spectrum of the missing data, which manifest as linear dipping co-phase axes in the $f\text{-}k$ domain. Except for SeisFusion*, SeisDDIMCR, and our method, all other methods suffer from severe spectral leakage and exhibit noticeable linear inclined artifacts.
\begin{table*}[!htbp]
\renewcommand{\arraystretch}{1.2} 
\setlength{\tabcolsep}{6pt} 
\caption{Comparison of various methods on the test set of different datasets with consecutive missing traces. The best results are highlighted in bold.}
\centering
\begin{tabular}{lrrrrrrrrrrrrrrrr}
\toprule 
\textbf{Dataset} & \multicolumn{4}{c}{SEG C3} & \multicolumn{4}{c}{MAVO} & \multicolumn{4}{c}{Model94}\\
Model & MSE$\downarrow$ & SNR$\uparrow$ & PSNR$\uparrow$ &SSIM$\uparrow$ & MSE$\downarrow$ & SNR$\uparrow$ & PSNR$\uparrow$ &SSIM$\uparrow$& MSE$\downarrow$ & SNR$\uparrow$ & PSNR$\uparrow$  &SSIM$\uparrow$\\
\midrule  
\textbf{DD-CGAN}\cite{Chang2020} & 7.545e-04 & 25.269  & 31.223 & 0.891 & 3.624e-04 & 28.877 & 34.409 & 0.938 & 1.683e-03 & 21.167 & 27.739 & 0.917\\ 
\textbf{cWGAN-GP}\cite{wei2022big} & 3.180e-04 & 29.022 & 34.976 & 0.935 & 2.153e-04 & 31.138 & 36.670 & 0.960 & 7.705e-04 & 24.560 & 31.132 & 0.951\\ 
\textbf{PConv-UNet}\cite{pan2020partial} & 4.023e-04 & 28.000 & 33.954 & 0.933 & 1.577e-04 & 32.490 & 38.022 & 0.972 & 9.343e-04 & 23.723 & 30.295 & 0.945\\  
\textbf{ANet}\cite{Yu9390348}& 4.428e-04 & 27.584 & 33.538 & 0.935 & 2.028e-04 & 31.397 & 36.929 & 0.965 & 9.207e-04 & 23.787 & 30.359 & 0.949\\ 
\textbf{Coarse-to-Fine}\cite{wei2022hybrid} & 2.244e-04 & 30.535 & 36.489 & 0.960 & 1.429e-04 & 32.918 & 38.450 & 0.972 & 7.585e-04 & 24.629 & 31.201 & 0.951\\ 
\textbf{SeisFusion*}\cite{10681481} & 3.982e-04 & 28.045 & 33.999 & 0.953 & 1.611e-04 & 32.398 & 37.930 & 0.973 & 1.082e-03 & 23.085 & 29.657 & 0.946\\ 
\textbf{SeisDDIMCR} \cite{10731876}& 1.601e-04 & 32.002 & 37.956 & 0.973 & 9.741e-05 & 34.582 & 40.114  & 0.979 & 5.913e-04 & 25.710 & 32.282 & 0.966 \\  
\textbf{Ours} & \bf{1.369e-04} & \bf{32.682} & \bf{38.636} & \bf{0.973} & \bf{9.234e-05} & \bf{34.814} & \bf{40.346} & \bf{0.979} & \bf{4.418e-04} & \bf{26.976} & \bf{33.548} & \bf{0.968}\\ 
\bottomrule
\end{tabular}
  \label{tab:continuousmissing}
  \vspace{-2mm}
\end{table*}
\begin{figure*}[!htbp]
  \centering
    \subfloat[Ground Truth.]
  {\includegraphics[height=0.145\textwidth]{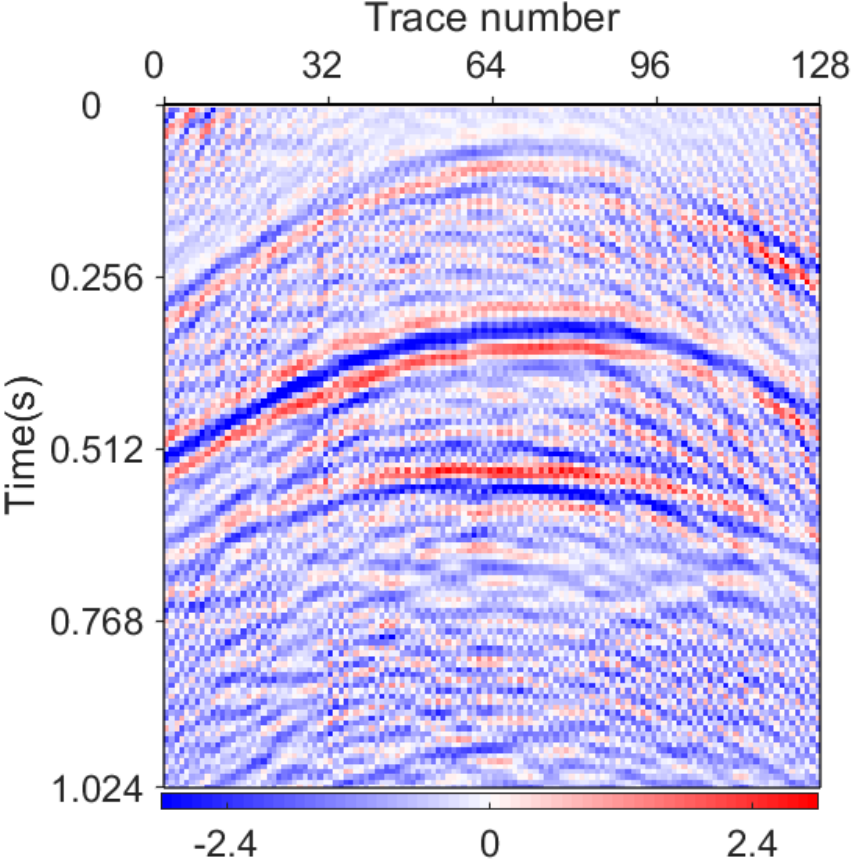}\label{fig:segc3continus_subfig1}\hspace{1.0mm}}
  \subfloat[DD-CGAN.]
  {\includegraphics[height=0.145\textwidth]{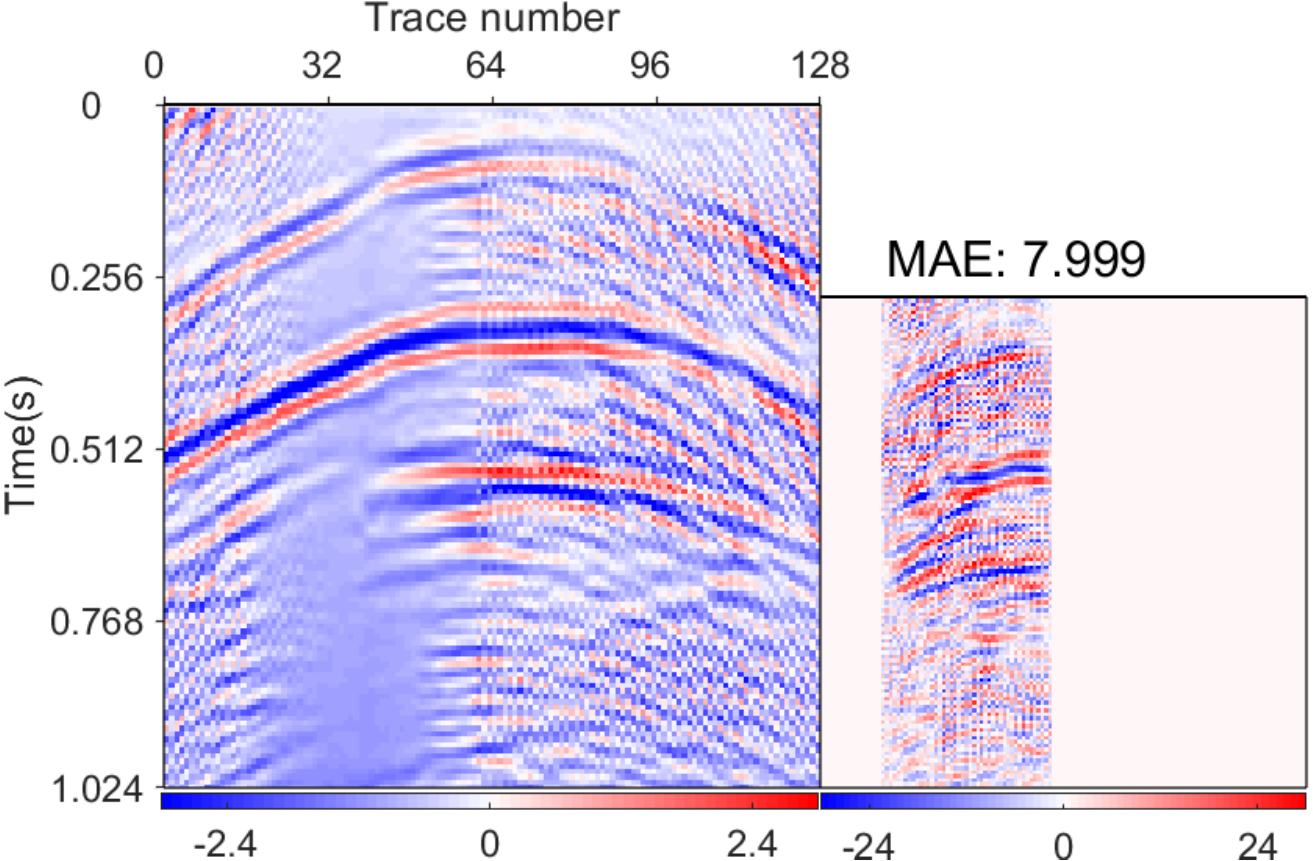}\label{fig:segc3continus_subfig3}}
  \subfloat[cWGAN-GP.]
  {\includegraphics[height=0.145\textwidth]{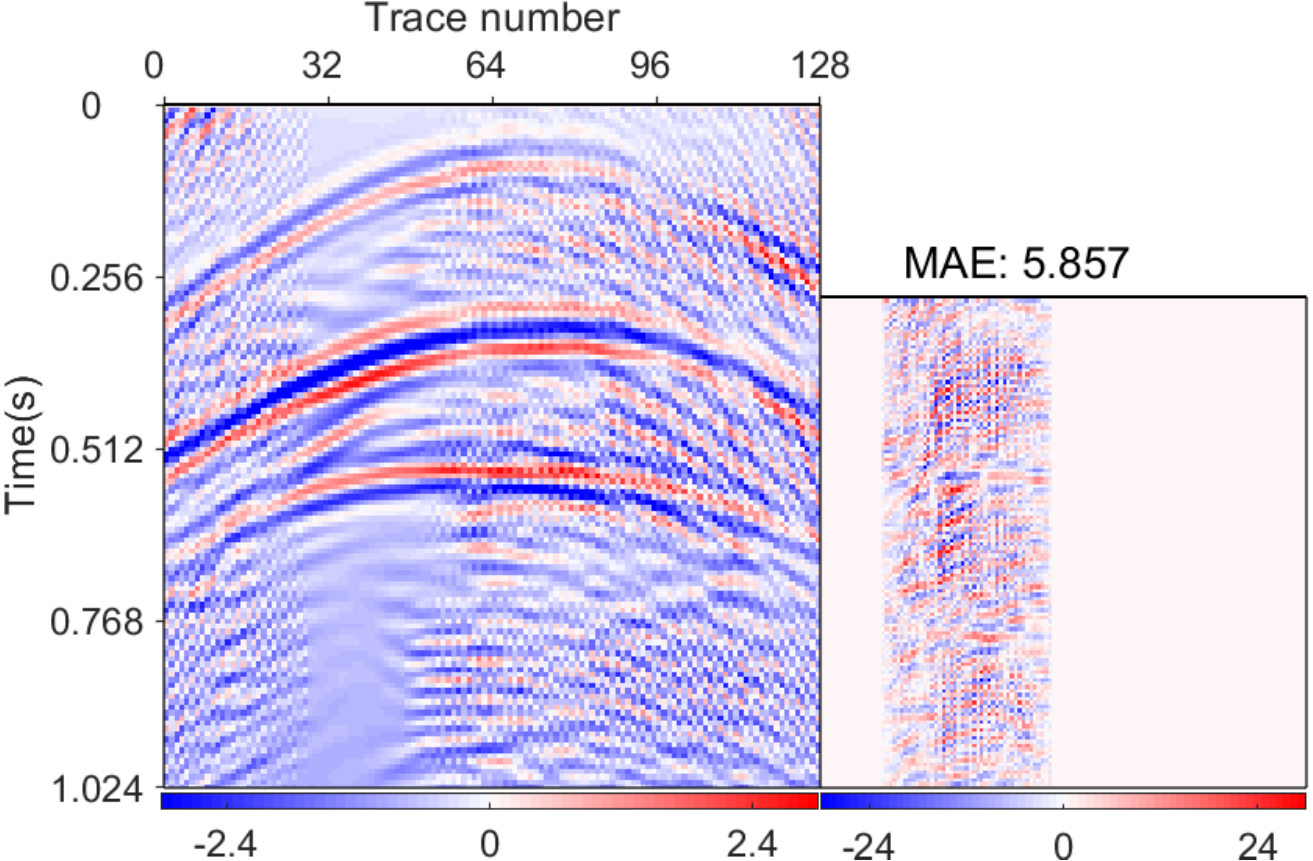}\label{fig:segc3continus_subfig4}}
  \subfloat[PConv-UNet.]
  {\includegraphics[height=0.145\textwidth]{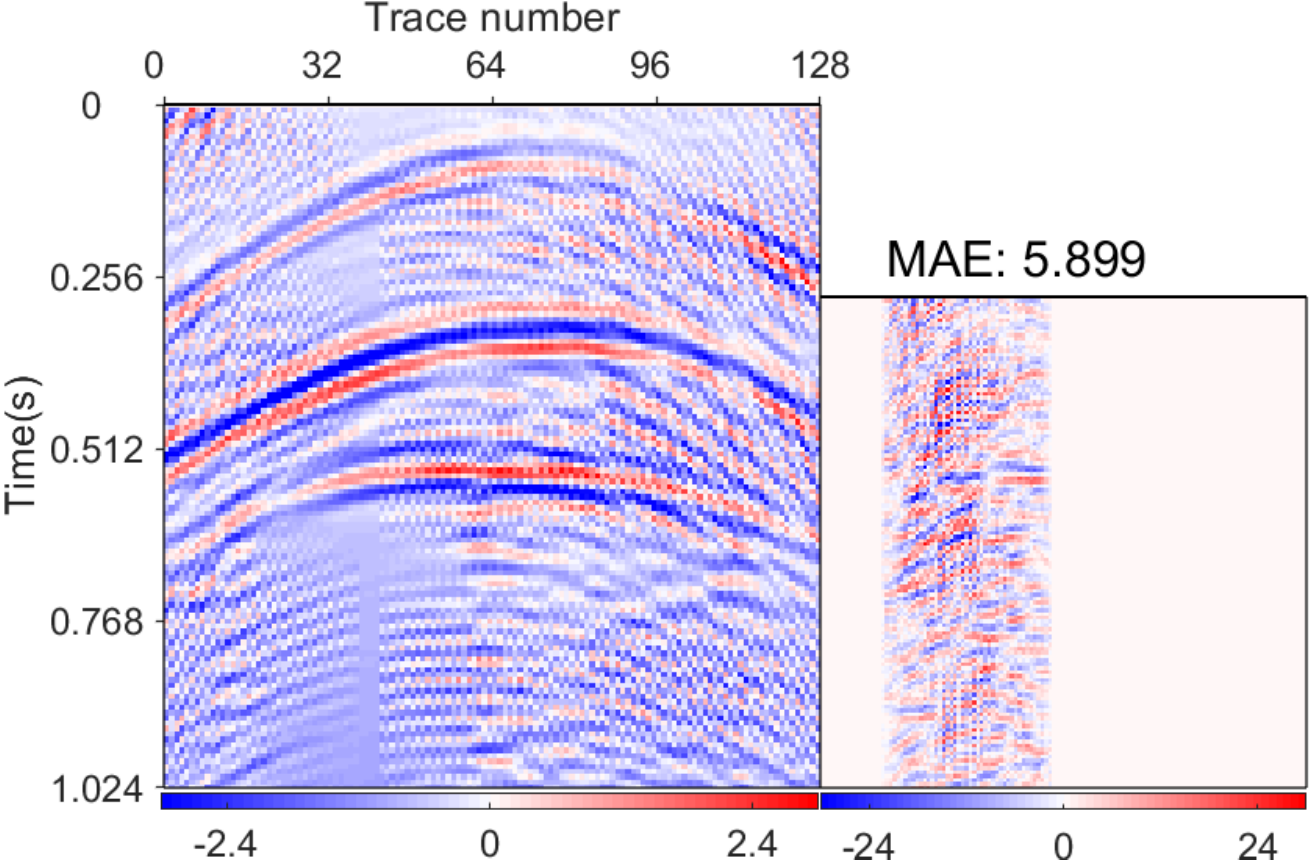}\label{fig:segc3continus_subfig5}}
  \subfloat[ANet.]
  {\includegraphics[height=0.145\textwidth]{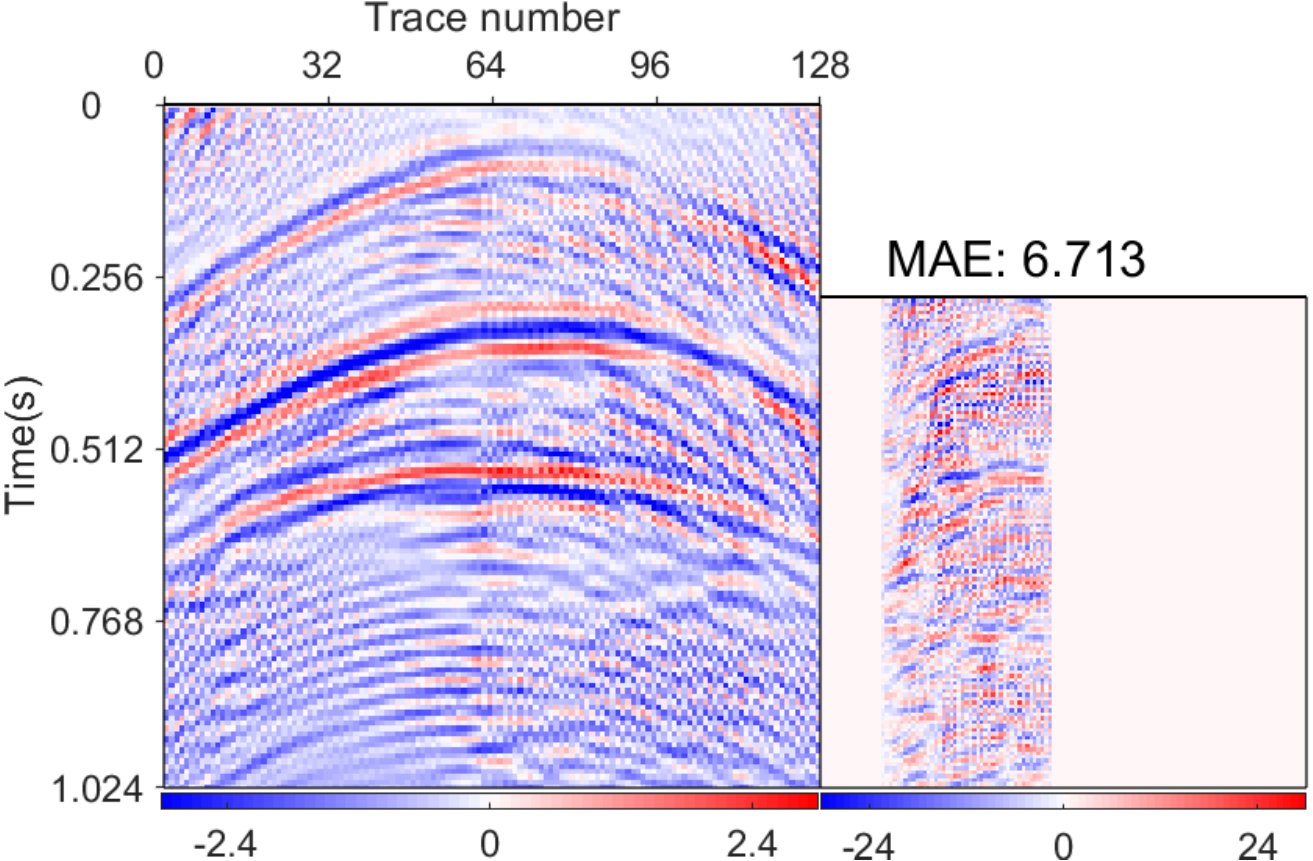}\label{fig:segc3continus_subfig6}}
 \clearpage
  \subfloat[Missing data.]
{\includegraphics[height=0.145\textwidth]{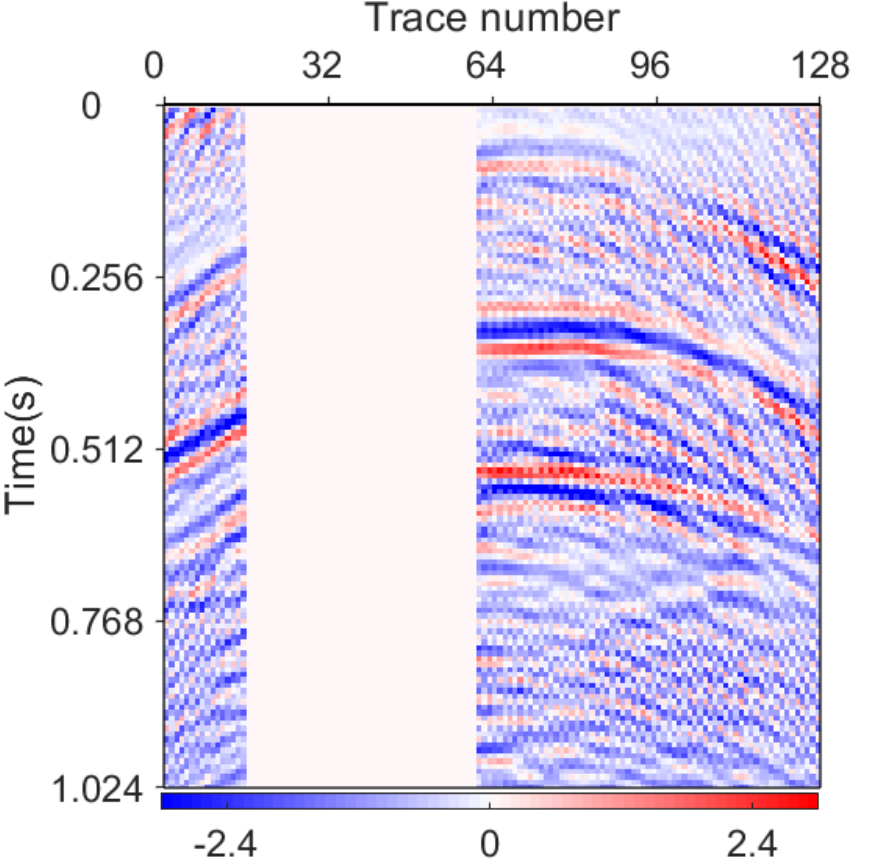}\label{fig:segc3continus_subfig2}}
  \subfloat[Coarse-to-Fine.]  {\includegraphics[height=0.145\textwidth]{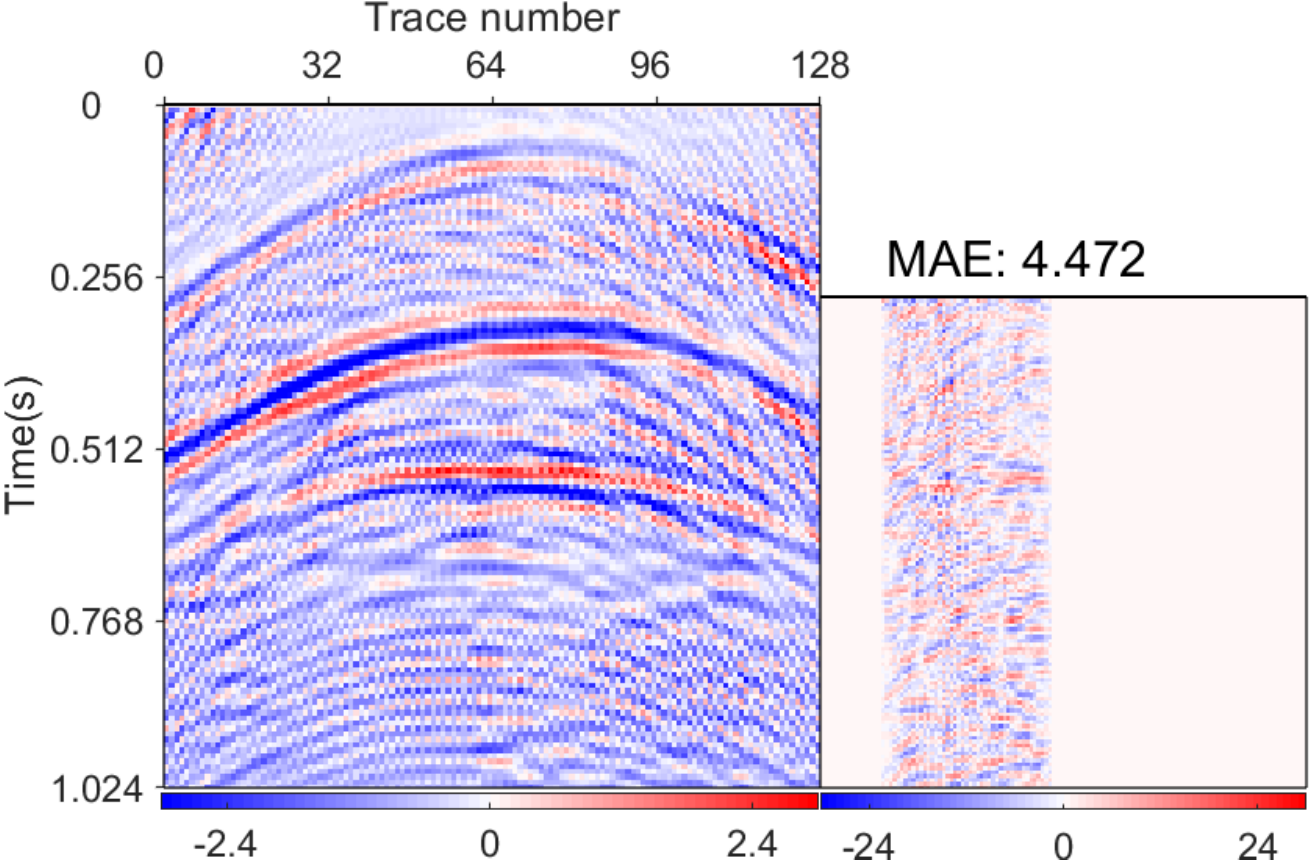}\label{fig:segc3continus_subfig7}}
  \subfloat[SeisFusion*.]
  {\includegraphics[height=0.145\textwidth]{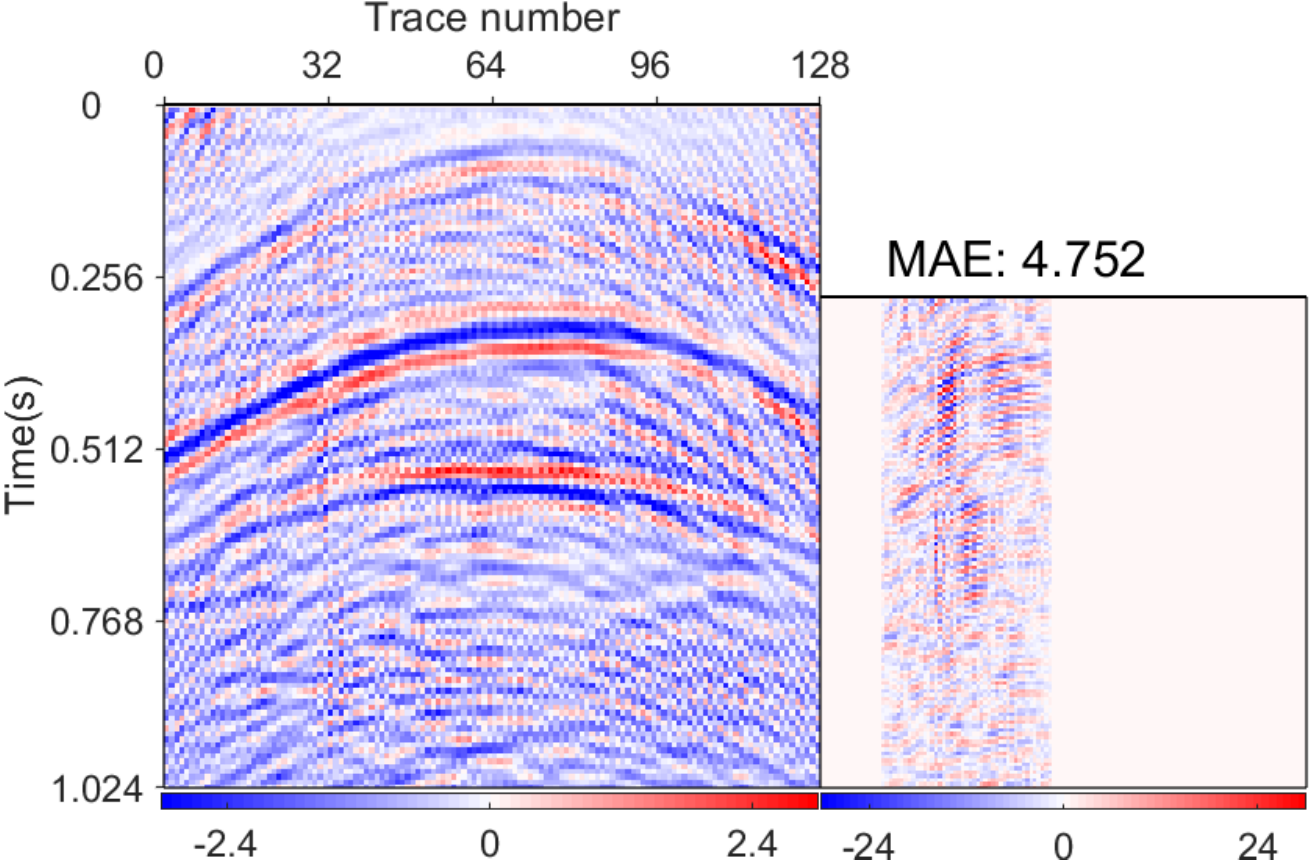}\label{fig:segc3continus_subfig8}}
  \subfloat[SeisDDIMCR.] 
    {\includegraphics[height=0.145\textwidth]{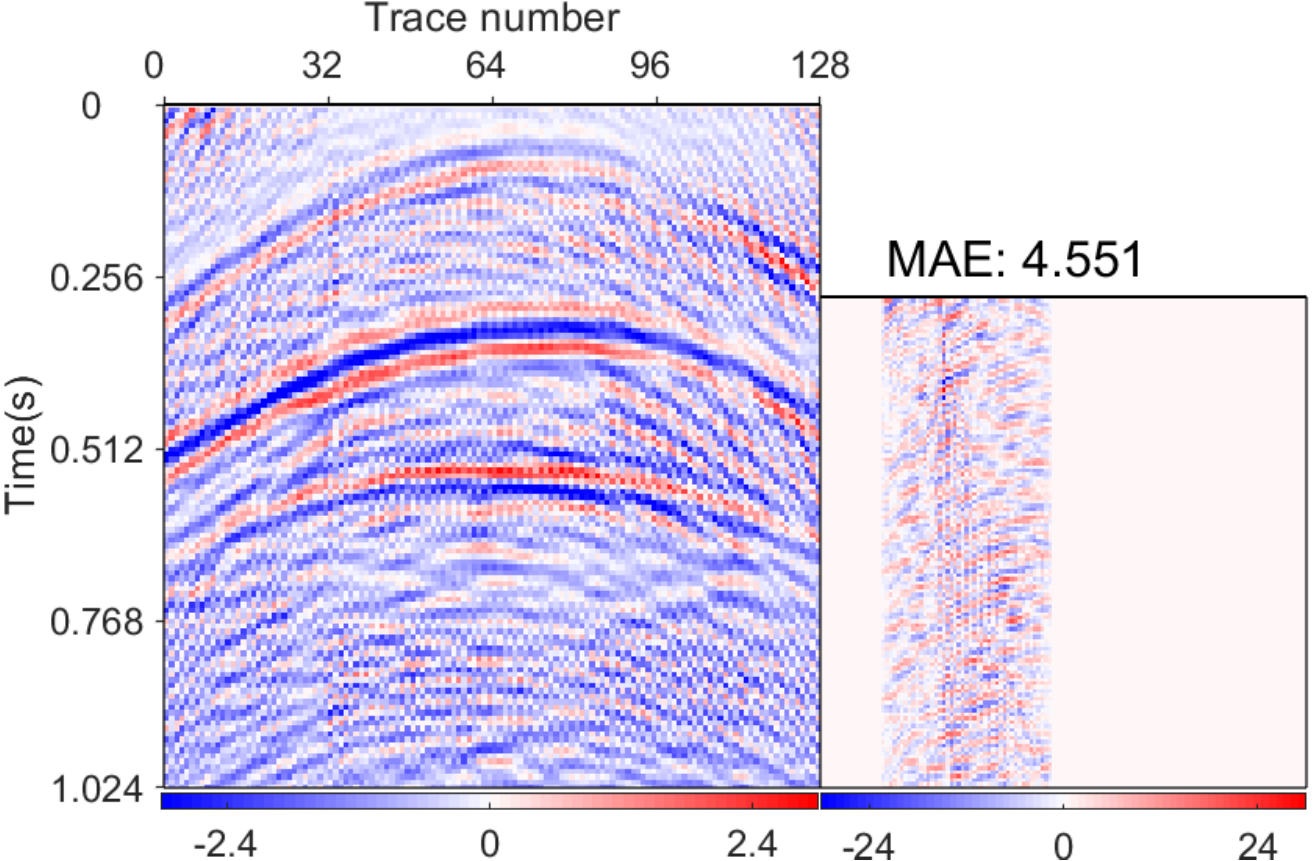}\label{fig:segc3continus_subfig9}}
  \subfloat[Ours.]
    {\includegraphics[height=0.145\textwidth]{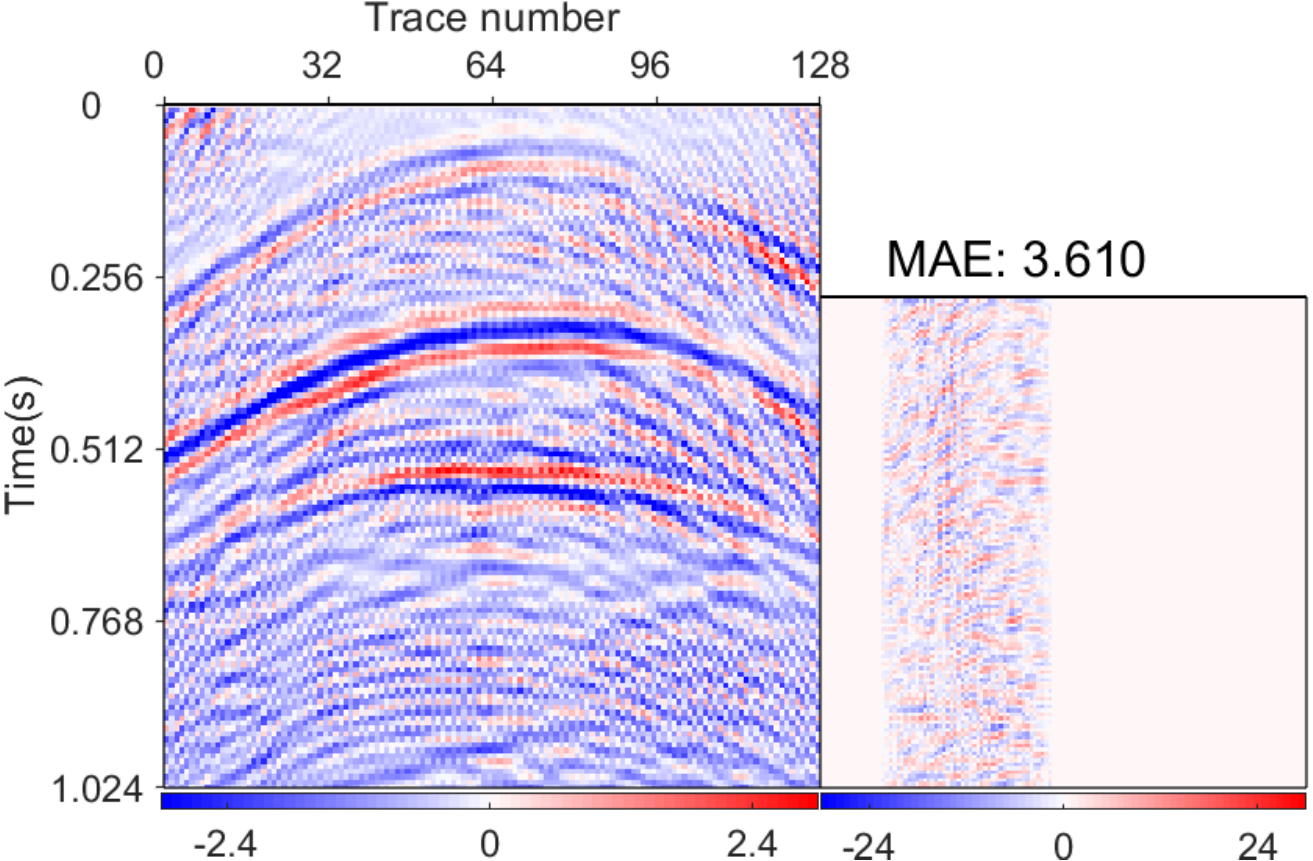}\label{fig:segc3continus_subfig10}}
  
  \caption{Interpolation results of the SEG C3 patch with a 35\% continuous missing ratio on different methods. We restore seismic data to its native amplitude range and apply the gain method to enhance subtle waveforms. Reconstruction residuals (right panel) are evaluated directly from raw data, excluding gain processing to preserve authentic signal differences. MAE is marked above the residual plot.}
  \label{fig:SEGC3 continus}
\vspace{-4mm}
\end{figure*}
The central energy concentration region (main signal) of our method is similar to other approaches, but it exhibits significantly lower noise in the low-frequency range (indicated by the black dashed box) and reduced energy leakage, benefiting from the precise reconstruction of randomly missing traces. Our method achieves a frequency distribution more consistent with the ground truth (e.g., within the solid black box), recovering relatively more mid- and high-frequency components.

\subsubsection{Continuous Missing Traces}\label{exp:Consecutive Missing Traces}
\begin{figure}[htbp]
    \centering
    \includegraphics[height=3.9cm]{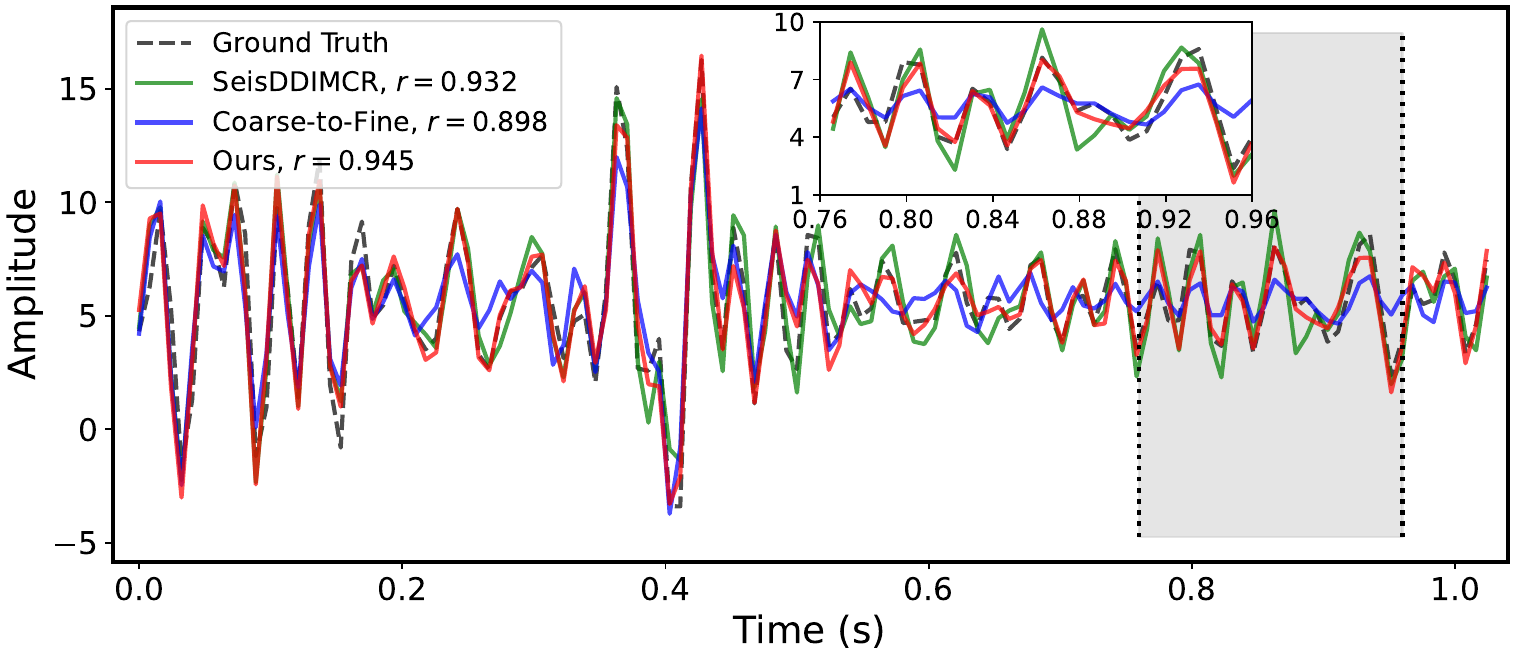}
    \vspace{-8mm}
    \caption{The wiggle plot of continuous missing gap interpolation results for a single SEG C3 test trace using different methods. We select a representative missing trace (marked by the black dashed line) to evaluate interpolation performance while quantitatively calculating the Pearson correlation coefficient $r$ between the reconstructed and true amplitudes.} 
    \label{fig:SEGC3 wiggle}  
\vspace{-2mm}
\end{figure}

The continuous missing rate in the test scenario remains between 0.1 and 0.6. Tab. \ref{tab:continuousmissing} presents the interpolation metrics on the three test sets with continuous missing gaps. Compared with other approaches, our method demonstrates superior performance in both fidelity preservation and textural detail retention. Fig. \ref{fig:SEGC3 continus} compares the interpolation differences between different methods by showing amplitude recovery results and residual maps for a local SEG C3 test patch with 35\% missing gaps. DD-CGAN, cWGAN-GP, PConv-UNet, and ANet fail significantly in signal recovery, with clear deviations in waveform continuity and amplitude fidelity. Blurred boundary information leads to unreasonably abrupt changes or artifacts. In contrast, methods like Coarse-to-Fine, SeisFusion*, and SeisDDIMCR show improvements, generating visually reasonable interpolations, yet still have issues such as local amplitude anomalies. Our method, with the smallest Mean Absolute Error (MAE), produces the most consistent recovery with the real signal. Fig. \ref{fig:SEGC3 wiggle} shows the amplitude wiggle plots of different methods for a single SEG C3 seismic trace with continuous missing gaps. Our method achieves the best global waveform matching, with the highest Pearson correlation coefficient ($r$$=$0.945). In the zoomed-in region, the amplitude variation of our method closely matches the ground truth. More recovery results for continuous missing interpolation in the $x\text{-}t$ domain and $f\text{-}k$ domain are provided in the supplementary material.

\begin{figure}[htbp]
    \centering
    \subfloat[]
    {\includegraphics[height=4.2cm]{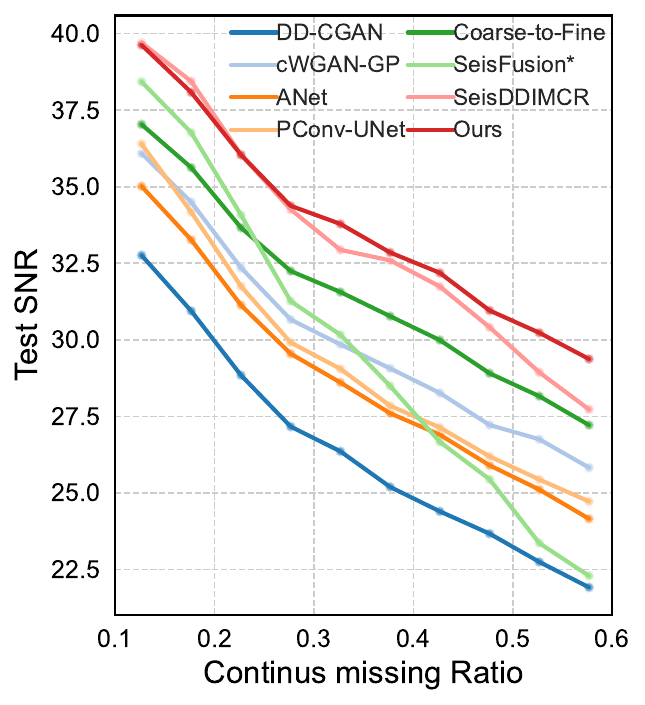}\label{fig:global_snr_binned_1}}
    \hspace{-2mm}
    \subfloat[]
    {\includegraphics[height=4.2cm]{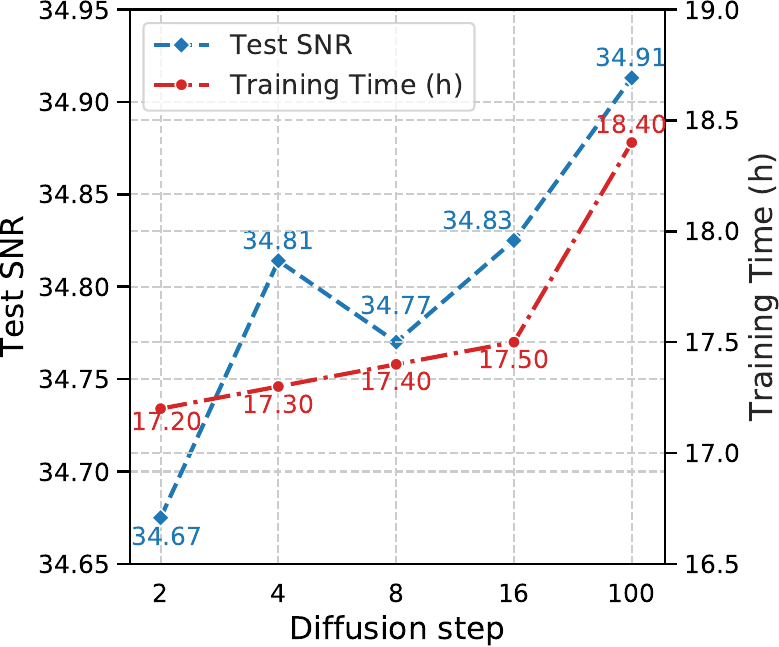}\label{fig:ablationdiffusion_steps}}
    \vspace{-1mm}
    \caption{(a) Curves of SNR versus the continuous missing ratio for different methods on the SEG C3 test dataset. (b) The comparison of test SNR and training time cost for our method under different diffusion steps on the MAVO dataset with continuous missing traces.} 
    \label{fig:global_snr_binned}  
\end{figure}

The continuous missing rate significantly impacts seismic trace interpolation performance. As systematically compared in Fig. \ref{fig:global_snr_binned}(a) using SNR curves of the SEG C3 test dataset, all methods exhibit SNR degradation with increasing missing rates, and demonstrate distinct attenuation gradients. While the two diffusion-based models and our method exhibit comparable performance at low missing rates ($<$20\%), a marked performance divergence emerges under high missing rates ($>$30\%), particularly highlighting our approach’s enhanced robustness in severe data-absence scenarios. Notably, SeisFusion* suffers drastic SNR deterioration ($\triangle \text{SNR}$$=$-16.1 dB from 10\% to 60\% missing rates), revealing that plug-and-play conditional resampling strategies relying solely on available data inevitably induce error accumulation issues, particularly exacerbated under severe missing conditions. Our method demonstrates consistently superior interpolation performance across all missing rates. 

\section{Ablation Study}\label{secabalation}
This section conducts systematic ablation studies to validate the efficacy of individual components (Spaformer and SPA modules) in the proposed methodology. Further complexity and computational efficiency analyses substantiate the superiority of our approach. Besides, we provide the ablation study on the loss function in the supplementary materials.


\subsection{Spaformer}\label{secabalation_spaformer}
The Spaformer constitutes the core of our framework as a U-Net-based end-to-end generator that integrates compressed seismic priors encoded through SPEN and generated via diffusion processes. Its architecture employs transformer blocks as fundamental units, incorporating multiscale feature construction through SPA modules and feature propagation or enhancement via FFN. Due to the structural complexity of SPA modules, their detailed analysis is reserved for Section \ref{secabalation_spa}. This section systematically examines the ablation effects of SPEN, diffusion processes, FFN, and the adaptive gating operation of the dynamic feature calibration mechanism (commonly used in FFN and SPA). Tab. \ref{tab:ablationSpaformer} quantifies the performance degradation observed on the MAVO test set with continuous missing traces when selectively disabling these components, revealing their distinct functional roles in maintaining waveform fidelity and texture generation capabilities. The first row of Tab. \ref{tab:ablationSpaformer} presents the benchmark performance metrics of our complete model. To systematically assess the SPEN module’s contributions, two ablation configurations are implemented. The first one is the non-shared SPEN architecture with independently initialized encoders for $\boldsymbol{X}_{obs}$ and $\boldsymbol{X}$ (Row 2 in Tab. \ref{tab:ablationSpaformer}). The second configuration completely removes SPEN, thereby eliminating both prior learning and the diffusion process (as shown in Row 3 of Tab. \ref{tab:ablationSpaformer}). Experimental results demonstrate hierarchical performance degradation. The non-shared configuration reduces test SNR by 0.16 dB due to impaired cross-path knowledge transfer, while complete SPEN removal causes more severe degradation of test SNR ($\triangle \text{SNR}=-0.33 \text{ dB}$). This quantitative analysis reveals that shared SPEN parameters enable inter-path feature correlation, while the integrated prior learning contributes greater fidelity enhancement than standalone U-Net architecture, establishing SPEN’s pivotal role in seismic signal reconstruction. The ablation study with diffusion process elimination (Row 4 in Tab. \ref{tab:ablationSpaformer}) causes performance degradation (0.32 dB SNR drop). Fig. \ref{fig:global_snr_binned}(b) illustrates the variation in model performance and training time with different diffusion steps on the MAVO dataset. This phenomenon demonstrates an inherent accuracy-efficiency trade-off, i.e., increasing diffusion steps from 2 to 100 steps progressively enhances test SNR (34.67→34.91 dB) yet linearly extends training duration (17.20→18.40 hours), necessitating systematic balancing in diffusion step configuration. The ablation study results in Row 5 of Tab. \ref{tab:ablationSpaformer} highlight the critical role of FFN, as its removal significantly weakens feature enhancement, evidenced by a 2.4 dB SNR drop.
The experimental results in Row 6 of Tab. \ref{tab:ablationSpaformer} show that the removal of the adaptive gating mechanism causes a subtle SNR reduction, demonstrating its fine-grained modulation in dynamic weight allocation.

\begin{table}[htbp]
\scriptsize
\setlength{\tabcolsep}{5.2pt}
\vspace{-3mm}
\caption{Ablation of different configurations of Spaformer. }
\centering
\vspace{-3mm}
\begin{tabular}{llllllll}
\toprule  
 \multirow{2}{*}{SPEN}& \multirow{2}{*}{Diffusion} & \multirow{2}{*}{FFN} & \multirow{2}{*}{Gate} & \multirow{2}{*}{MSE} & \multirow{2}{*}{SNR}  & \multirow{2}{*}{PSNR} & \multirow{2}{*}{SSIM}\\ %
\\[-2.0mm] 
& Process &  & & & & \\
\midrule  
 \checkmark & \checkmark  & \checkmark & \checkmark & \bf{9.234e-05} & \bf{34.814} & \bf{40.346} & \bf{0.979} \\
  \midrule 
Non-shared &  & && 9.532e-05 &34.659 &40.208&0.979\\
\ding{55} & \ding{55} & & & 9.952e-05 &34.489 &40.021&0.978\\
 & \ding{55}&  & & 9.944e-05 &34.492 &40.024&0.978\\
 &  & \ding{55} & & 1.606e-04 &32.410 &37.942&0.969\\
 &  &  & \ding{55} & 9.434e-05 & 34.721 & 40.253 & 0.979 \\ 
\bottomrule
\end{tabular}
  \label{tab:ablationSpaformer}
\vspace{-6mm}
\end{table}


\begin{figure*}[!htbp]
\centering
\subfloat[]{
    \includegraphics[height=5.5cm] {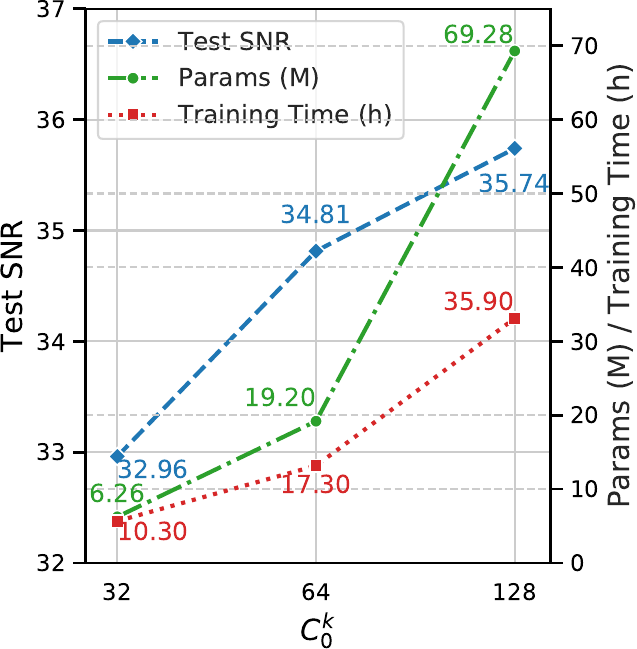}
    \label{fig:complexity_1}
}
\subfloat[]{
    \includegraphics[height=5.5cm]{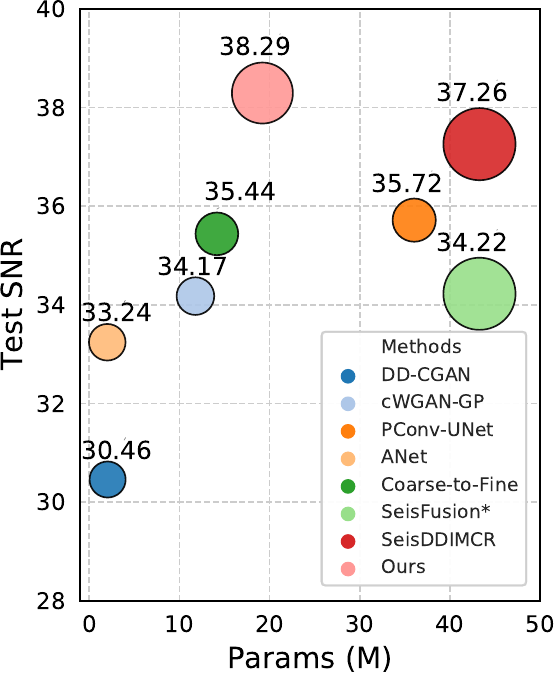}
    \label{fig:complexity_2}
}
\subfloat[]{
    \includegraphics[height=5.5cm] {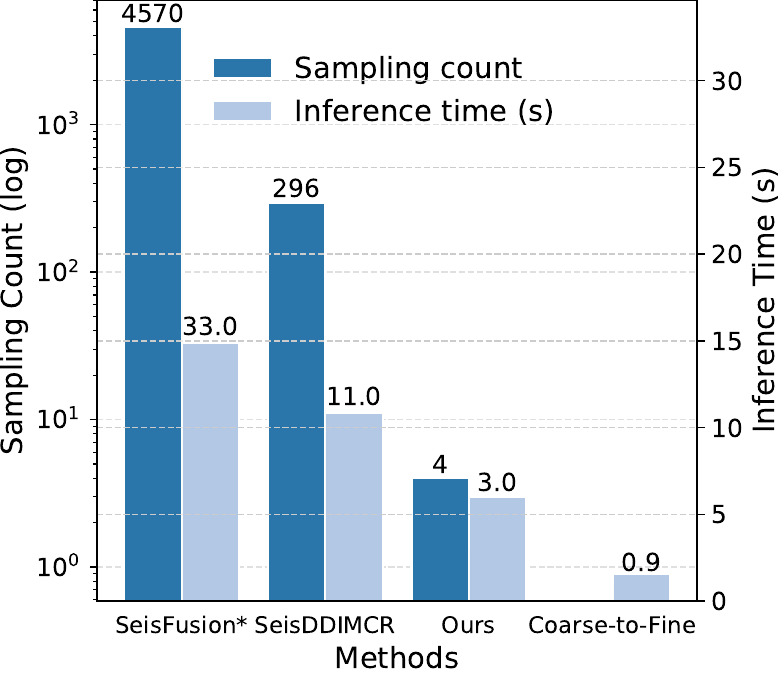}
    \label{fig:complexity_3}
}
\vspace{-2mm}
\caption{(a) The comparison of the test SNR, number of parameters (Params), and training time cost for our method under different selections of $C^k_0$ on the MAVO dataset with continuous missing traces. 
(b) The comparison of Params and test SNR for different methods on the SEG C3 dataset with random missing traces. The bubble size scales proportionally to the FLOPs, reflecting the computational cost of a model. 
(c) The comparison of the inference time (evaluated on a single-sample basis) and sampling count for different methods on the MAVO dataset with continuous missing traces. The sampling count axis is shown on a log scale.}
\label{fig:complexity}
\end{figure*}

\subsection{SPA}\label{secabalation_spa}
SPA serves as the core module for establishing similarity metrics and building global correlations. To validate the effectiveness of our proposed sparse attention calculation method based on L2 similarity, we compare the model's performance and calculation efficiency under different similarity 
measurement methods and key space dimensions ($C^k_0$, defined by the first-layer channel dimension of the U-Net). The channel dimensions of the four layers are $C^k_0$, $2C^k_0$, $4C^k_0$, and $8C^k_0$, respectively. The results are summarized in Tab. \ref{tab:ablationSPA}. Increasing $C^k_0$ significantly reduces reconstruction error and improves quality, as higher dimensions capture more detailed structural information. However, as Fig. \ref{fig:complexity}(a) shows, the performance gain diminishes with increasing dimensionality, with the improvement from 64 to 128 dimensions smaller than from 32 to 64. Both the total model parameters (Params (M)) and training cost exhibit superlinear growth with the increase of $C^k_0$. Additionally, it can also be observed that FLOPs grow nonlinearly with dimension, while the size of keys increases linearly in Tab. \ref{tab:ablationSPA}.  Therefore, we adopt the moderate dimensionality of 64 to balance the performance and computational efficiency. Ablation experiments show that with equal feature dimensionality, L2 similarity consistently outperforms cosine similarity, and occupies similar computational (FLOPs) and memory costs (size of keys). This may be due to L2's sensitivity to feature magnitude, which aligns with the physical meaning of amplitude errors in seismic data, enabling more accurate matching without extra cost.

\begin{table}[h]
\scriptsize
\begin{threeparttable}
\setlength{\tabcolsep}{3pt}
 \caption{Ablation of different configurations of SPA.}
  \centering
  \vspace{-3mm}
\begin{tabular}{llllllll}
\toprule
\multirow{2}{*}{Similarity}  & \multirow{2}{*}{$C^k_0$} &\multirow{2}{*}{MSE} & \multirow{2}{*}{SNR}  & \multirow{2}{*}{PSNR} & \multirow{2}{*}{SSIM} & \multirow{2}{*}{FLOPs(G)} & \multirow{2}{*}{Size of }\\ 
\\[-2.0mm] 
function& && & & & &keys(MB)\\ 
\midrule
L2 similarity  & 64 &9.234e-05 & 34.814 & 40.346 & 0.979  &6.26 &3.41 \\ 
Cosine similarity & 64 & 9.794e-05  & 34.559 & 40.091 &0.979 & 6.23 &3.41\\

 \midrule
 L2 similarity  & 32 & 1.414e-04  & 32.962 & 38.494 &0.972& 1.57 & 0.85\\ 
Cosine similarity  & 32 & 1.486e-04  & 32.746 & 38.278 &0.971 & 1.56 & 0.85\\ 
 \midrule
 L2 similarity  & 128 & 7.462e-05  & 35.740 & 41.272 &0.983 & 24.97 & 13.63\\ 
Cosine similarity  & 128 & 7.569e-05  & 35.677 & 41.209 &0.983 & 24.90 & 13.63\\ 
\bottomrule
\end{tabular}
\label{tab:ablationSPA}
\begin{tablenotes}
\scriptsize
\item[*] The number of floating-point operations (FLOPs) is calculated solely for the similarity operation.
\end{tablenotes}
\end{threeparttable}
  \vspace{-6mm}
\end{table}

\subsection{Computational Complexity}\label{secabalation_ComputationalComplexity}
We evaluate our model from three aspects, i.e., parameter size, computational cost, and inference cost. As shown in Fig. \ref{fig:complexity}(b), the joint visualization of the computational complexity and interpolation performance of random missing traces on the SEG C3 dataset reveals differences in the accuracy-complexity trade-offs on various models. Model sizes range from 2M (ANet and DD-CGAN) to 43M (SeisDDIMCR). Our method achieves the highest SNR with a moderate model size of 19M, outperforming the second-best SeisDDIMCR by 1.03 dB, thereby demonstrating superior efficiency and performance. Fig. \ref{fig:complexity}(c) compares the inference efficiency of three diffusion architectures (SeisFusion*, SeisDDIMCR, and Ours) with the end-to-end baseline (Coarse-to-Fine) on the MAVO dataset. SeisFusion* and SeisDDIMCR suffer from heavy resampling overhead, limiting real-time processing capability. By leveraging transformer-based feature extraction and SPEN-guided prior projection, our method reduces the sampling process to just four steps, achieving an inference speed comparable to that of the end-to-end baseline.

\section{Conclusion}\label{secConclusion}
In this paper, we propose Diff-spaformer, a novel framework that integrates transformer architecture with diffusion processes to address the challenge of seismic data interpolation. SPEN effectively bridges the global modeling capability of sparse multi-head attention with the distribution consistency constraints of diffusion models. We demonstrate the superior performance of L2 similarity over traditional cosine similarity in seismic amplitude modeling. Sparse self-attention and adaptive ReLU filtering significantly reduce computational complexity while maintaining feature interaction efficiency. The framework employs a single-stage optimization and deterministic lightweight sampling strategy, improving both efficiency and interpolation quality. Experimental results show strong performance in handling random and continuous missing data, avoiding the computational burden of iterative resampling in plug-and-play diffusion interpolation models, with promising potential for field data applications. Future work will focus on 3D seismic data interpolation and domain generalization strategies to enhance robustness across exploration scenarios. 
\section*{Acknowledgment}

The authors thank Sandia National Laboratory, Mobil Oil Company, and Dr. Joe Dellinger at BP for their provision of open-access seismic data resources that facilitate the development and validation of the proposed methodology throughout this investigation.


\ifCLASSOPTIONcaptionsoff
  \newpage
\fi

\small
\bibliographystyle{IEEEtran}
\bibliography{reference}

\end{document}